\documentclass{jfm}
\usepackage{graphicx}
\usepackage{epstopdf, epsfig}
\usepackage{amsmath}
\usepackage{tikz,pgfplots}
\usepackage{tikz-3dplot}
\pgfplotsset{compat=1.10}
\usetikzlibrary{decorations.pathreplacing}
\usepackage{physics}
\usepackage{graphicx}
\usepackage{caption}
\usepackage{subcaption}
\usepackage[normalem]{ulem}
\usepackage{xcolor}

\newcommand\redsout{\bgroup\markoverwith{\textcolor{red}{\rule[0.5ex]{2pt}{0.4pt}}}\ULon}
\newcommand\Redsout{\bgroup\markoverwith{\textcolor{red}{\rule[1.2ex]{2pt}{0.4pt}}}\ULon}

\def\BV{Brunt-V{\"a}is{\"a}l{\"a}\ }

\shorttitle{Mean flow generation by three-dimensional non-linear internal wave beams}
\shortauthor{F. Beckebanze and L. R. M. Maas}

\shortauthor{F. Beckebanze,
  K. J. Raja, and L. R. M. Maas}

\title{Mean flow generation by three-dimensional non-linear internal wave beams}

\author{F. Beckebanze\aff{1}
  \corresp{\email{f.beckebanze@uu.nl}},
  K. J. Raja\aff{2},  
  \and L. R. M. Maas\aff{3}}

\affiliation{\aff{1} Mathematical Institute, Utrecht University, P.O. Box 80010, 3508 TA Utrecht, The Netherlands
\aff{2} Laboratoire des \'{E}coulements G\'{e}ophysiques et Industriels, Universit\'{e} Grenoble Alpes,
Grenoble, CS 40700, France
\aff{3}Institute for Marine and Atmospheric research Utrecht (IMAU), Utrecht University, Princetonplein 5, 3584 CC Utrecht, The Netherlands}

\begin{document}

\maketitle
\begin{abstract}
We study the generation of resonantly growing mean flow by weakly non-linear internal wave beams. With a perturbational expansion, we construct analytic solutions for 3D internal wave beams, exact up to first order accuracy in the viscosity parameter. We specifically focus on the subtleties of wave beam generation by oscillating boundaries, such as wave makers in laboratory set-ups. The exact solutions to the linearized equations allow us to derive an analytic expression for the mean vertical vorticity production term, which induces a horizontal mean flow. Whereas mean flow generation associated with viscous beam attenuation - known as streaming - has been described before, we are the first to also include a peculiar inviscid mean flow generation in the vicinity of the oscillating wall, resulting from line vortices at the lateral edges of the oscillating boundary. Our theoretical expression for the mean vertical vorticity production is in good agreement with earlier laboratory experiments, for which the previously unrecognized inviscid mean flow generation mechanism turns out to be significant.
\end{abstract}


\section{Introduction}
Internal waves are ubiquitous in stratified and/or rotating fluids, such as the oceans. Typical occurrence of internal waves includes oblique beams that propagate at a fixed angle with respect to the horizontal, the angle $\theta = \arctan  \sqrt{ \frac{\omega_0^2-f_0^2}{N_0^2-\omega_0^2} }$ being controlled by the wave frequency $\omega_0$, the natural buoyancy frequency of the ambient stratified fluid $N_0$, and the Coriolis frequency $f_0$. It has been recognized that internal waves play an important role in mixing the abyssal oceans \citep{WF04} and marginal seas \citep{La14}. Whereas the primary generation mechanism of internal waves through tidal conversion at rough topography is fairly well understood \citep{GK07}, it is still debated as to which mechanisms dissipate the internal waves \citep{SS02, Da17}. ~ \\  
An important dissipation mechanism of internal waves can be the generation of mean flow \citep{Bu10}, and in particular horizontal mean flow associated with mean vertical vorticity, here referred to as \emph{vortical} induced mean flow, sometimes called \emph{strong} mean flow \citep{Bo12, Da17}. 
The hallmark of the potentially strong \emph{vortical} induced mean flow is the persistent, cumulative transfer of energy from the wave field. By contrast, the typically weak \emph{buoyancy advection}-induced mean flow \citep{KC01, TA03}, comprising the induced mean \emph{horizontal} vorticity, is strongly suppressed by the background stratification. The buoyancy advection-induced mean flow vanishes in the absence of viscosity except where internal wave beams intersect \citep{Th87, TAL05}. We emphasize that the buoyancy advection-induced mean flow is absent in inviscid internal wave packages, which are instead accompanied by the well-studied modulation-induced mean flow \citep{Br69, TA07}, also known as Bretherton flow \citep{BS18}. Modulations of the internal wave field - not considered in this body of work - may contribute to the vortical induced mean flow \citep{KA15}.   \\ 
A prominent class of underlying mean flow generation mechanisms for internal wave beams is the so-called streaming \citep{Li78}, which entails mean flow generation associated with viscous attenuation through non-linear wave-wave interaction, similar to streaming by acoustic waves, and analogous to mean flow generation by surface waves \citep{LH64}. As reviewed by \cite{Ri01}, streaming also occurs in a large variety of homogeneous fluid configurations. Key ingredients for mean vertical vorticity production through streaming by internal waves beams - and hence \emph{vortical} mean flow generation - are both viscous attenuation and horizontal-cross-beam variations of the wave beam amplitude. Several recent studies investigate vortical mean flow generation through streaming in truly three-dimensional settings, both numerically \citep{KZS10, GB12, vdB14, ZD15, Ra18} and experimentally \citep{Gr10, Bo12, Gr13, Se16, Ka17}. 
\\
Over long time scales, the vortical induced mean flow may become sufficiently energetic such that wave-mean flow interactions eventually lead to a breakdown of the internal wave itself. This breakdown mechanism is referred to as streaming instability if the underlying generation mechanism is associated with irreversible energy conversion from the wave to the mean field \citep{Da17}. This differs from self-acceleration, which refers to inviscid modulation-induced mean flow advecting the waves until they become convectively unstable \citep{Su06}. 
\\
An approximate expression for the mean vertical vorticity production through streaming for monochromatic internal wave beams was derived by \cite{Bo12}, and extended to slowly time-varying wave beams by \cite{KA15} and nearly monochromatic wave packages by \cite{FKA18}. Their analyses rely on scale separation in the along-beam velocity, $u$, and the horizontal cross-beam velocity, $v$, which limits the applicability of their asymptotic results. We find that this scale separation is not always justified in the vicinity of an oscillating boundary. \\
In this study, we construct analytical solutions for 3D internal wave beams, exact up to first order accuracy in the viscosity parameter, generated by oscillating boundaries, such as wave makers \citep{Go07}. The velocity field satisfying the linearized equations also includes purely horizontal wave motion associated with vertical line-vortices at the edges of the wave maker. Our analytic expression for the mean vertical vorticity production term includes the well-known streaming as well as a peculiar inviscid mean flow generation in the neighborhood of the oscillating boundary associated with the vertical line-vortices. The relative strengths of the line vortices - and hence the relative importance of the associated mean flow generation - strongly depends on the mathematical representation of the oscillating boundary. For this reason, we present a detailed derivation of an appropriate mathematical representation of a small-amplitude wall oscillation, which differs from popular representations in numerical simulations. Our analysis suggests that streaming and inviscid mean flow generation by the line vortices are equally important in energizing the vortical induced mean flow in the laboratory experiments by \cite{Bo12}. Our theory cannot describe the long-term mean flow evolution as we ignore the feedback of the growing mean flow on the beam evolution.
\\
The paper is organized as follows. We first present preliminaries in \S \ref{preliminaries} and derive an appropriate mathematical idealization of small-amplitude boundary oscillation in \S \ref{energy_source}. Analytical expressions for monochromatic 3D internal wave beams solving the linearized equations for the oscillating boundary representation are constructed in \S \ref{3D_sol}, and used in \S \ref{streaming} to determine the associated mean vertical vorticity production. \S \ref{comparison} is devoted to a thorough comparison of our new theoretical insights with the laboratory experiments by \cite{Bo12}.
A discussion of our results, as well as suggestions for insightful analysis of experimental internal wave field data, can be found in \S \ref{conclusions}.
\section{Preliminaries}
\label{preliminaries}
We consider a uniformly-stratified incompressible Boussinesq fluid with \BV frequency $N_0>0$ on an $f$-plane in Cartesian coordinates $(x,y,z)$, rotating around the vertical axis $z$ with half the Coriolis frequency $f_0>0$, and where gravity points along the negative $z$-direction. In such rotating stratified fluids, inviscid internal waves with frequencies $\omega_0>0$ propagate at angle $\theta=\arctan  \sqrt{\frac{\omega_0^2-f_0^2}{N_0^2-\omega_0^2} }$ with respect to the horizontal if either $f_0<\omega_0<N_0$ or $N_0<\omega_0<f_0$. For localized energy sources oscillating at frequency $\omega_0$, this leads in 2D to the well-known St. Andrew Cross (e.g. \cite{Su10}, also sketched in Fig. \ref{sketch_energy_input}(a)) and in 3D to double cones \citep{Vo03}. Throughout this article, we consider boundary forcings at a vertical sheet, oscillating at frequency $\omega_0$, representative of small-amplitude wall oscillations, such as sketched in Fig. \ref{sketch_energy_input}(b). The precise oscillating boundary formulation is described in \S \ref{energy_source}. \\We shall work with dimensionless variables, employing some characteristic wave length $L_0$ as length scale, $1/\omega_0$ as the time scale, and $U_0 = a_0 \omega_0 = \epsilon L_0 \omega_0$ as the velocity scale, where $a_0$ is the dimensional wave amplitude and $\epsilon = a_0/L_0 \ll 1$ is the Stokes number.  \\
The governing equations for the non-dimensional velocity vector ${\bf u}=[u,v,w]$, the buoyancy $b$ and the pressure $p$ with dimensionless \BV frequency $N=N_0/\omega_0$ and Coriolis frequency $f=f_0/\omega_0$ in subscript-derivative notation are
\begin{align}
     {\bf u}_t+\epsilon \left( {\bf u} \cdot  \nabla \right) {\bf u}   + f \hat{z} \wedge {\bf u}  & =   - \nabla p+\vartheta \Delta {\bf u}+\hat{z}b,     \label{ge1_mom}      \\ 
    b_t +\epsilon \ {\bf u}  \cdot  \nabla b                                       				         &  =  -N^2 w,         \label{ge1_buoy}       \\
     \nabla  \cdot {\bf u}     							    					 & =        0.       \label{ge1_cont}
\end{align}
Here, $\vartheta=\frac{\nu}{\omega_0 L_0^2}$, with $\nu$ the kinematic viscosity constant, and $\vartheta/\epsilon$ is the inverse Reynolds number. We assume both $\epsilon$ and $\vartheta$ to be small, allowing us to perform a perturbational expansion in these parameters. \\
We proceed by employing the wave/vortex decomposition for the horizontal flow \citep{SR89, Vo03}. The Helmholtz decomposition thus reads
\begin{equation}
{\bf u} = \nabla_h \phi   +  \hat{z} \wedge \nabla \Psi + \hat{z} w,
\label{wv_dec}
\end{equation}
where $\phi$ is the potential (wave), and $\Psi$ is the stream function (vortex) of the horizontal velocity. The usefulness of this decomposition relies on the absence of vertical vorticity ($\Omega^z = \Delta_h \Psi$) for internal gravity wave fields, i.e. the vortex stream function $\Psi$ is \emph{not} associated with internal wave motion. The decomposition (\ref{wv_dec}) transforms the continuity equation (\ref{ge1_cont}) into
\begin{equation}
\Delta_h \phi  + w_z = 0.
\label{cont_Phi_w}
\end{equation}
The lower index $h$ denotes horizontal components only, i.e. $\nabla_h=[\partial_x,\partial_y,0]$, $\Delta_h = \nabla_h^2$. \\
The curl of the horizontal momentum equations in (\ref{ge1_mom}), $\hat{z} \wedge \nabla$, reduces to
\begin{equation}
\left( \partial_t-\vartheta \Delta \right) \Delta_h \Psi + f \Delta_h \phi =  \  \epsilon R  ,
\label{curl_mom}
\end{equation}
where
\begin{equation}
\begin{split}
  R  			=   &     \left(  ({\bf u} \cdot \nabla ) u         \right)_y   -   \left(  ({\bf u} \cdot \nabla )v         \right)_x \\
  			=   &  	\underbrace{ J(w, \phi_z) 	}_{\mbox{\tiny wave - wave} }
			      + 	\underbrace{ J(\Psi, \Delta_h \Psi) }_{\mbox{\tiny vortex-vortex} }  +
	\underbrace{  w_z  \Delta_h \Psi  	-   \left( \nabla_h   w  \right) \cdot \left( \nabla_h   \Psi_z   \right)  	
	- 			\left( \nabla_h  \phi   \right) \cdot \left( \nabla_h  \Delta_h \Psi  \right) 	-    w  \Delta_h \Psi_z   }_{\mbox{ \tiny wave - vortex }},
\label{eq_R}
\end{split}
\end{equation}
and $J(A,B) = A_x B_y    -   B_x A_y $ is the horizontal Jacobian. The vertical curl of the non-linear horizontal advection terms, $R$, constituting the mean vertical vorticity production upon time-averaging over the wave period, is analyzed in detail in \S \ref{streaming}. \\
Similarly, the divergence of the horizontal momentum equations becomes
\begin{equation}
\left( \partial_t-\vartheta \Delta \right)  \Delta_h \phi - f \Delta_h \Psi = -\Delta_h p    \ + \ \mathcal{O}(\epsilon).
\label{div_horiz_mom}
\end{equation}
This allows us to relate the pressure $p$ to the horizontal wave and vortex components, $\phi$ and $\Psi$ respectively, through
\begin{equation}
p=-\left( \partial_t-\vartheta \Delta \right) \phi   +   f \Psi    + \Psi^{cf} +   \mathcal{O}(\epsilon),
\nonumber
\end{equation}
where curl-free stream function $\Psi^{cf}$ is a harmonic gauge, satisfying $\Delta_h \Psi^{cf} =0$, and determined by appropriate boundary conditions. \\
Expressing the  vertical momentum equation in terms of $w$ and $\phi$ by employing the buoyancy equation (\ref{ge1_buoy}), Eq. (\ref{curl_mom}) and Eq. (\ref{div_horiz_mom}) gives
\begin{equation}  
\begin{split}
	\left[ \left( \partial_t-\vartheta \Delta \right)^2    +   f^2 \right] & \partial_t \partial_z \Delta_h \phi      		-		     \left[   \left( \partial_t-\vartheta \Delta \right) \partial_t + N^2  \right] \left( \partial_t-\vartheta \Delta \right) \Delta_h w 	=  0 \   +		\ 				\mathcal{O}(\epsilon), 		
\end{split}
\label{eq_Phi_w}
\end{equation}
Using the continuity equation (\ref{cont_Phi_w}), we can reduce (\ref{eq_Phi_w}) to
\begin{equation}
 \left[ \left( \partial_t-\vartheta \Delta \right)^2   \partial_t \Delta       	+ 		f^2 \partial_{t} \partial_z^2 		+		N^2 (\partial_t - \vartheta \Delta) \Delta_h \right] w		= 0  \   +		\ \mathcal{O}(\epsilon).
 \label{eq_w}
\end{equation}

Next, we discuss and specify boundary constraints representative of small-amplitude wall oscillations, for which we derive analytic solutions to the equations (\ref{eq_w}) and (\ref{curl_mom}) up to $\mathcal{O}(\vartheta, \epsilon^0)$-accuracy in \S \ref{3D_sol}.

\section{Mathematical representation of oscillating boundary forcing}
\label{energy_source}

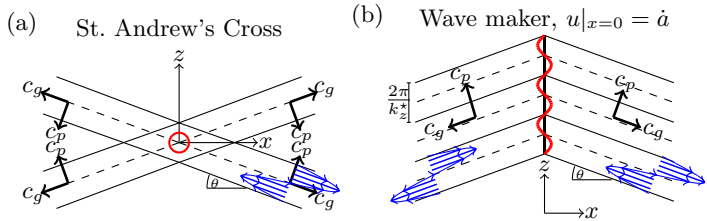
\begin{figure}
\begin{center}
\hspace*{-1.0cm}
\begin{minipage}{.33\textwidth}
 \begin{tikzpicture}[scale=0.26]
\draw[->] (0,0) -- (0,4);
\node at ( 0 , + 4.4) {$z$};
\draw[->] ( 0 , 0  ) -- ( 4 , 0  );
\node at ( 4.4 , 0  ) {$x$};

\draw[-,black,thin] ({  sin(70)*4 + cos(70)*(-1)}, { (sin(70))*(-1) - cos(70)*4} )-- ({ -2+  sin(70)*4 + cos(70)*(-1)}, { (sin(70))*(-1) - cos(70)*4} ); 
\node[black] at   ({ -1.6+  sin(70)*4 + cos(70)*(-1)}, { 0.25 + (sin(70))*(-1) - cos(70)*4} )  {\tiny $\theta$};
\draw [thin,domain=0:20] plot  ({  sin(70)*4 + cos(70)*(-1)  - 2*cos(\x)}, { (sin(70))*(-1) - cos(70)*4   + 2 * sin(\x)} ) ;

\draw[red,thick] (0, 0) circle (0.5 cm);

\draw[->,black,line width=0.35mm] ({ sin(70)*6 + cos(70)*0}, { (sin(70))*0 - cos(70)*6}) -- ({  sin(70)*6 + cos(70)*1.5}, { (sin(70))*1.5 - cos(70)*6} );
\node[black] at ({  sin(70)*6 + cos(70)*1.8}, { (sin(70))*1.8 - cos(70)*6} ) {$c_p$}; 
\draw[->,black,line width=0.35mm] ({ sin(70)*6 - cos(70)*0}, { (sin(70))*0 + cos(70)*6}) -- ({  sin(70)*6 - cos(70)*(-1.5)}, { (sin(70))*(-1.5) + cos(70)*6} );
\node[black] at ({  sin(70)*6 - cos(70)*(-1.8)}, { (sin(70))*(-1.8) + cos(70)*6} ) {$c_p$}; 
\draw[->,black,line width=0.35mm] ({ - sin(70)*6 + cos(70)*0}, { -(sin(70))*0 - cos(70)*6}) -- ({ - sin(70)*6 + cos(70)*(-1.5)}, { -(sin(70))*(-1.5) - cos(70)*6} );
\node[black] at (-{  sin(70)*6 + cos(70)*(-1.8)}, { -(sin(70))*(-1.8) - cos(70)*6} ) {$c_p$}; 
\draw[->,black,line width=0.35mm] ({- sin(70)*6 - cos(70)*0}, {- (sin(70))*0 + cos(70)*6}) -- ({-  sin(70)*6 - cos(70)*(1.5)}, { -(sin(70))*(1.5) + cos(70)*6} );
\node[black] at ({ - sin(70)*6 - cos(70)*(1.8)}, { -(sin(70))*(1.8) + cos(70)*6} ) {$c_p$}; 

\draw[->,black,line width=0.35mm] ({ sin(70)*6 + cos(70)*0}, { (sin(70))*0 - cos(70)*6}) -- ({  sin(70)*7.5 + cos(70)*0}, { (sin(70))*0 - cos(70)*7.5} );
\node[black] at ({  sin(70)*7.9 + cos(70)*0}, { (sin(70))*0 - cos(70)*7.9 -0.3} ) {$c_g$}; 
\draw[->,black,line width=0.35mm] ({ sin(70)*6 - cos(70)*0}, { (sin(70))*0 + cos(70)*6}) -- ({  sin(70)*7.5 - cos(70)*0)}, { (sin(70))*0 + cos(70)*7.5} );
\node[black] at ({  sin(70)*7.9 - cos(70)*0}, { (sin(70))*0 + cos(70)*7.9} ) {$c_g$}; 
\draw[->,black,line width=0.35mm] ({ - sin(70)*6 + cos(70)*0}, { -(sin(70))*0 - cos(70)*6}) -- ({ - sin(70)*7.5 + cos(70)*0)}, { -(sin(70))*0 - cos(70)*7.5} );
\node[black] at (-{  sin(70)*7.9 + cos(70)*0}, { -(sin(70))*0 - cos(70)*7.9} ) {$c_g$}; 
\draw[->,black,line width=0.35mm] ({- sin(70)*6 - cos(70)*0}, {- (sin(70))*0 + cos(70)*6}) -- ({-  sin(70)*7.5 - cos(70)*0}, { -(sin(70))*0 + cos(70)*7.5} );
\node[black] at ({ - sin(70)*7.9 - cos(70)*0}, { -(sin(70))*0 + cos(70)*7.9} ) {$c_g$}; 

\foreach \d in {-1, 1}  
{\draw[-,black] (0, \d) -- ({  sin(70)*7 + cos(70)*\d}, { sin(70)*\d - cos(70)*7} );
}
\draw[-,dashed,black] (0, 0) -- ({  sin(70)*7 + cos(70)*0}, { (sin(70))*0 - cos(70)*7} );
\foreach \d in {-1, 1}  
{\draw[-,black] (0, \d) -- ({  sin(70)*7 - cos(70)*\d}, { sin(70)*\d + cos(70)*7} ); }
\draw[-,dashed,black] (0, 0) -- ({  sin(70)*7 - cos(70)*0}, { (sin(70))*0 + cos(70)*7} ); 
\foreach \d in {-1, 1}  
{\draw[-,black] (0, \d) -- ({  -sin(70)*7 - cos(70)*\d}, { sin(70)*\d - cos(70)*7} );
}
\draw[-,dashed,black] (0, 0) -- ({  -sin(70)*7 - cos(70)*0}, { (sin(70))*0 - cos(70)*7} );
\foreach \d in {-1, 1}  
{\draw[-,black] (0, \d) -- ({  -sin(70)*7 + cos(70)*\d}, { sin(70)*\d + cos(70)*7} );
}
\draw[-,dashed,black] (0, 0) -- ({  -sin(70)*7 + cos(70)*0}, { (sin(70))*0 + cos(70)*7} );

\foreach \d in {0, 0.1, 0.2, 0.3, 0.4, 0.5, 0.6, 0.7, 0.8, 0.9, 1.0, 1.1, 1.2, 1.3, 1.4, 1.5, 1.6, 1.7, 1.8, 1.9} 
{ \draw[-,blue] ({ 5.3+ cos(70)*\d - sin(70)*2.5*sin(180*(\d   )    )}, { -3 + sin(70)*\d + cos(70)*2.5*sin(180*(\d ))})  -- ({ 5.3+ cos(70)*(\d+0.1) - sin(70)*2.5*sin(180*(\d  +0.1) )}, { -3+ sin(70)*(\d+0.1) + cos(70)*2.5*sin(180*(\d  +0.1) )} );
}
\foreach \d in { 0.25,  	0.5	, 0.75,  1.01, 1.25, 	1.5,	1.75} 
{ \draw[->,blue,line width=0.20mm] ({5.3 + cos(70)*\d }, { -3 + sin(70)*\d } ) --    ({ 5.3+ cos(70)*\d - sin(70)*2.5*sin(180*(\d+0))}, { -3+ sin(70)*\d + cos(70)*2.5*sin(180*(\d+0))}) ;
}
\node[black] at (0, 5.8) {\footnotesize St. Andrew's Cross};

\node[black] at (-8, 6) { (a) };

\end{tikzpicture}
\end{minipage}
\begin{minipage}{.33\textwidth}
 \begin{tikzpicture}[scale=0.26]
\draw[->] (0,-4) -- (0,-2);
\node at ( 0 , -4 + 2.4) {$z$};
\draw[->] ( 0 , 0 - 4 ) -- ( 2 , 0 - 4 );
\node at ( 2.4 , 0 - 4 ) {$x$};

\draw[-,black,thin] ({  sin(70)*4 + cos(70)*(-1)}, { (sin(70))*(-1) - cos(70)*4} )-- ({ -2+  sin(70)*4 + cos(70)*(-1)}, { (sin(70))*(-1) - cos(70)*4} ); 
\node[black] at   ({ -1.6+  sin(70)*4 + cos(70)*(-1)}, { 0.25 + (sin(70))*(-1) - cos(70)*4} )  {\tiny $\theta$};
\draw [thin,domain=0:20] plot  ({  sin(70)*4 + cos(70)*(-1)  - 2*cos(\x)}, { (sin(70))*(-1) - cos(70)*4   + 2 * sin(\x)} ) ;

\draw[->,black,line width=0.35mm] ({ sin(70)*3 + cos(70)*2}, { (sin(70))*2 - cos(70)*3}) -- ({  sin(70)*3 + cos(70)*3.5}, { (sin(70))*3.5 - cos(70)*3} );
\node[black] at ({  sin(70)*3 + cos(70)*3.8}, { (sin(70))*3.8 - cos(70)*3} ) {$c_p$}; 
\draw[->,black,line width=0.35mm] ({ - sin(70)*3 - cos(70)*2}, { (sin(70))*2 - cos(70)*3}) -- ({ - sin(70)*3 + cos(70)*(-3.5)}, { -(sin(70))*(-3.5) - cos(70)*2} );
\node[black] at (-{  sin(70)*3 + cos(70)*(-4.2)}, { -(sin(70))*(-4.2) - cos(70)*3} ) {$c_p$}; 

\draw[->,black,line width=0.35mm] ({ sin(70)*3 + cos(70)*2}, { (sin(70))*2 - cos(70)*3}) -- ({  sin(70)*4.5 + cos(70)*2}, { (sin(70))*2 - cos(70)*4.5} );
\node[black] at ({  sin(70)*5.2 + cos(70)*2}, { (sin(70))*2 - cos(70)*5.2} ) {$c_g$}; 
\draw[->,black,line width=0.35mm] ({ - sin(70)*3 - cos(70)*2}, { (sin(70))*2 - cos(70)*3}) -- ({ - sin(70)*4.5 - cos(70)*2)}, { (sin(70))*2 - cos(70)*4.5} );
\node[black] at ({  -sin(70)*5.2 - cos(70)*2}, { (sin(70))*2 - cos(70)*5.2} ) {$c_g$}; 

\foreach \d in {-1, 1, 3, 5}  
{\draw[-,black] (0, \d) -- ({  sin(70)*7 }, { \d - cos(70)*7} ); }
\foreach \d in {0, 2, 4}  
{\draw[-,dashed,black] (0, \d) -- ({  sin(70)*7 }, { \d - cos(70)*7} ); }
\foreach \d in {-1, 1, 3, 5}  
{\draw[-,black] (0, \d) -- ({  -sin(70)*7 }, { \d - cos(70)*7} ); }
\foreach \d in {0, 2, 4}  
{\draw[-,dashed,black] (0, \d) -- ({  -sin(70)*7 }, { \d - cos(70)*7} ); }

\foreach \d in {0, 0.1, 0.2, 0.3, 0.4, 0.5, 0.6, 0.7, 0.8, 0.9, 1.0, 1.1, 1.2, 1.3, 1.4, 1.5, 1.6, 1.7, 1.8, 1.9} 
{ \draw[-,blue] ({ 5.3+ cos(70)*\d - sin(70)*2.5*sin(180*(\d   )    )}, { -3 + sin(70)*\d + cos(70)*2.5*sin(180*(\d ))})  -- ({ 5.3+ cos(70)*(\d+0.1) - sin(70)*2.5*sin(180*(\d  +0.1) )}, { -3+ sin(70)*(\d+0.1) + cos(70)*2.5*sin(180*(\d  +0.1) )} );
\draw[-,blue] ({ -5.3 - cos(70)*\d - sin(70)*2.5*sin(180*(\d   )    )}, { -3 + sin(70)*\d - cos(70)*2.5*sin(180*(\d ))})  -- ({ -5.3 - cos(70)*(\d+0.1) - sin(70)*2.5*sin(180*(\d  +0.1) )}, { -3+ sin(70)*(\d+0.1) - cos(70)*2.5*sin(180*(\d  +0.1) )} );
}
\foreach \d in { 0.25,  	0.5	, 0.75,  1.01, 1.25, 	1.5,	1.75} 
{ 
\draw[->,blue,line width=0.20mm] ({5.3 + cos(70)*\d }, { -3 + sin(70)*\d } ) --    ({ 5.3+ cos(70)*\d - sin(70)*2.5*sin(180*(\d+0))}, { -3+ sin(70)*\d + cos(70)*2.5*sin(180*(\d+0))}) ; 
\draw[->,blue,line width=0.20mm] ({-5.3 - cos(70)*\d }, { -3 + sin(70)*\d } ) --    ({ - 5.3 - cos(70)*\d - sin(70)*2.5*sin(180*(\d+0))}, { -3+ sin(70)*\d - cos(70)*2.5*sin(180*(\d+0))}) ;
}

\draw[-,line width = 0.4 mm] (0,-1) -- (0,5);

\foreach \d in {0, 0.1, 0.2, 0.3, 0.4, 0.5, 0.6, 0.7, 0.8, 0.9, 1.0, 1.1, 1.2, 1.3, 1.4, 1.5, 1.6, 1.7, 1.8, 1.9, 2.0, 2.1, 2.2, 2.3, 2.4, 2.5, 2.6, 2.7, 2.8, 2.9, 3.0, 3.1, 3.2, 3.3, 3.4, 3.5, 3.6, 3.7, 3.8, 3.9, 4.0, 4.1, 4.2, 4.3, 4.4, 4.5, 4.6, 4.7, 4.8, 4.9, 5.0, 5.1, 5.2, 5.3, 5.4, 5.5, 5.6, 5.7, 5.8, 5.9}
{ \draw[-,red,line width = 0.4 mm] ({  0.35*(+ sin(180*\d)) }, {  \d -1 })  -- ( {0.35* (+ sin(180*(\d+0.1)))}, { \d-1 +0+0.1}) ;  }

\node[black] at (0, 5.8) {\footnotesize Wave maker, $u|_{x=0}=\dot{a}$};

\draw[-,thin, black] (-6.9     , 0.6   ) --  (-6.9      ,  2.6 ); 
\draw[-,thin, black] (-6.9+0.1, 0.6 ) --  (-6.9-0.1,  0.6) ;
\draw[-,thin, black] (-6.9+0.1, 2.6 ) --  (-6.9-0.1,  2.6) ;
\node at (-1.4 - 6, 1.6) {$\frac{2\pi}{k_z^{\star}}$};

\node[black] at (-9, 6) { (b) };

\end{tikzpicture}
\end{minipage}
\caption{(a)  Schematic snapshot of a St. Andrew's Cross generated by vertical oscillation of a line force (point source in $(x,z)$-plane); $c_p$ and $c_g$ indicating phase and group velocities. (b) Two wave beams generated by horizontal wave maker oscillations. We impose the horizontal velocity, $\dot{a}$, of the oscillating wave maker (red line) at the center line, $x=0$ (thick black line).     }
\label{sketch_energy_input}
\end{center}
\end{figure}

The aim is to formulate an appropriate mathematical description of a boundary value problem which is
(i) representative of a small-amplitude horizontally oscillating boundary, such as a wave maker in the laboratory set-up by \cite{Bo12}, and
(ii) suitable to solve the linearized equations (\ref{ge1_mom}) - (\ref{ge1_cont}) analytically. Reasonable simplifications are necessary because it is notoriously difficult to compute the wave field with velocity vector ${\bf u}$ satisfying the impermeability boundary condition at the oscillating wall, 
\begin{equation}
			\frac{d }{dt} \left( x-\epsilon a(t,y,z)\right) = 0  \qquad  \Leftrightarrow \qquad 			 		 {\bf n} \cdot {\bf u}|_{x=\epsilon a(y,z,t) } =  \dot{a}(y,z,t)   \label{constraint_u} 
\end{equation}
where $d/dt = \partial_t + \epsilon \ {\bf u} \cdot \nabla$ is the material derivative, and ${\bf n} = (1, -\epsilon \partial_y a, -\epsilon \partial_z a)$ is a vector normal to the oscillating boundary, $x=\epsilon a(y,z,t)$. \\ 
For computational convenience we restrict our analysis to temporally monochromatic wall oscillations, with non-dimensional frequency $\omega=1$\footnote{For the ease of dimensionalizing our expressions, we denote the frequency $1$ by $\omega$ everywhere. For dimensional expressions, simply replace $\omega$ by $\omega_0$ and $\vartheta$ by $\nu$ everywhere.}.  A straightforward generalization of our results to almost-periodic wave fields follows by the theory developed in \cite{Kr91}. We consider the phase propagation ($c_p$) of the oscillating wall to be primarily upward, such that the group velocity ($c_g$) of the generated wave field is primarily downward, as sketched in Fig. \ref{sketch_energy_input}b. Purely upward phase propagation is atypical for laboratory experiments; the relative strength of the upward propagating field component for primarily downward propagating beams is discussed in the appendix, \S \ref{slit_exp}.  

\noindent A key simplification, valid for all small-amplitude oscillations ($\epsilon \ll 1$), consists of prescribing the impermeability constraint (\ref{constraint_u})  at $x=0$ instead of $x=\epsilon a(y,z,t)$. If additionally, the forcing in the vertical sheet is spatially smooth ($\partial_y a, \partial_z a \in \mathcal{O}(1)$), then the constraint (\ref{constraint_u}) at $\mathcal{O}(\epsilon^{0})$-accuracy reduces to
\begin{equation}
u|_{x=0} = \dot{a}(y,z,t) .
\label{constraint_u2}
\end{equation}
The physical wave makers in laboratory set-ups, which we want to mimic mathematically, typically have sharp edges, where $\partial_y a \gg 1$, possibly $\partial_y a \geq 1/\epsilon $. This means that the constraint (\ref{constraint_u}) cannot be simplified to (\ref{constraint_u2}), i.e. the product of $\epsilon ( \partial_y a) $ and $ v|_{x=0}$ in (\ref{constraint_u}) is non-negligible. Recall that the monochromatic wall motion, $\dot{a}$, generates a wave field which oscillates predominately at frequency $\omega=1$. This means that the product of $\partial_y a$ with the dominating first harmonic of $v|_{x=0}$ (both oscillating at frequency $\omega$) results in a mean and/or second harmonic, which cannot be balanced by $\dot{a}$ or the first harmonic of ${\bf u}$. An unphysical blow-up of mean and/or second harmonics at the lateral edges of the wave maker is only circumvented if  $v|_{x=0}=0$ where $\partial_y a \gg 1$, and we are forced to add this condition as a constraint. While the impermeability constraint (\ref{constraint_u}) \emph{cannot} be reduced to constraint (\ref{constraint_u2}), we \emph{can} reduce it to
\begin{equation}
u|_{x=0} = \dot{a}(y,z,t) 		\qquad \mbox{everywhere, and } \qquad 		 v|_{x=0} = 0		  \qquad		\mbox{where}  \ \partial_y a \gg 1 .
\label{constraint_u3}
\end{equation}
We remark that similarly where $\partial_z a $ blows up, the vertical velocity should vanish ($w|_{x=0}=0$). Treatment of this additional boundary forcing constraint is neglected because it is irrelevant for the main objective of the present analysis\footnote{Large $\epsilon \partial_z a$ corresponds to a horizontal line source, generating a St. Andrew's Cross, as sketched in Fig. \ref{sketch_energy_input}(a), and discussed in \S \ref{comparison} and \S \ref{slit_exp}. No separate treatment of $\epsilon \partial_z a \gg 1$ is required because the present analysis, focusing on the \emph{vortical} induced mean flow generation, is unaltered by the weak upward-propagating branch of the St. Andrew Cross.}. We emphasize that the constraint on $v$ in (\ref{constraint_u3}) belongs to the impermeability constraint, i.e. it does \emph{not} specify a stress (no-slip) boundary condition for the along-wall velocity. This is important, because it implies that even though $v$ points along the wall $x=0$ in our approximate description, we may not use a Stokes boundary layer solution for $v$ to satisfy (\ref{constraint_u3}). \\

\noindent Let us first discuss two common approaches to implement (\ref{constraint_u2}), before we discuss our implementation of  (\ref{constraint_u3}). \\
The first approach consists in prescribing constraint (\ref{constraint_u}) or a related formulation. This approach is popular in simulations because its numerical implementation is straightforward. For an idealized wave maker with no-stress boundary conditions, Raja et al (2018) prescribe $u(x=0)= \dot{a}(t)$. Using no-slip boundaries, \cite{Br16} prescribe $w(x=0) =  -\dot{a}(t) \tan \theta$, which is equivalent to $u(x=0)= \dot{a}(t)$ for two-dimensional inviscid wave beams that propagate under an inclination $\tan \theta$. These representations of a wave maker generate an accurate internal wave far-field, i.e. at sufficient distance from the energy input. However, our analysis in \S \ref{3D_sol} reveals that these implementations fail to describe the wave field components related to vertical line vortices, which are inevitably generated at the lateral edges of the wave maker. Those line vortices do play a significant role in the mean vertical vorticity production, and are thus essential for our analysis. \\
The second common approach consists in prescribing a momentum body forcing ${\bf F} = \hat{x}\ddot{a}(y,z,t) \delta(x)$, where $\delta$ is the Dirac Delta. This approach is popular in theoretical studies because Green's functions of the governing equations are often known. For numerical implementations of this momentum-forcing-approach, the Dirac Delta is typically replaced by a sharp Gaussian. Although one can easily find equivalent body forcing formulations for forced boundary constraints for two-dimensional problems, it appears non-trivial to do so for three-dimensional problems. \\
We first solve the traditional boundary value constraint (\ref{constraint_u2}) approximately in terms of the wave potential ($\phi$), such that we can subsequently construct a stream function $\Psi$ to satisfy the additional constraint in (\ref{constraint_u3}) exactly. Although we are primarily interested in the half-open domain, $x>0$, we naturally extend the analytical expressions to $\mathbb{R}^3$ by taking $u$ and $v$ even in $x$, and $w$ odd in $x$, as sketched in Fig. \ref{sketch_energy_input}b. 

\subsection{Wave maker representation}
Throughout the remainder of this study, we consider 
$$  a(y,z,t) = - E(y,z) e^{-i \omega t} \qquad \mbox{with} \qquad E(y,z) =\frac{1 }{4\pi^2} \int \limits_{-\infty}^{\infty} \int \limits_0^{\infty} \hat{E}(k_y,k_z) e^{i k_y y + i k_z z} d k_z d k_y, $$
where the Fourier spectrum $\hat{E}$ of the normalized wall oscillation envelope is assumed to be negligible for all negative vertical wave numbers $k_z$ to guarantee upward phase propagation, an assumption that can be dropped whenever needed. We work with complex expressions; physical quantities always correspond to the real part. For wave makers of height $2l_z$, width $2l_y$, and vertical wave number $k_z^{\star}$, we take
\begin{equation}
\begin{split}
&  E(y,z) 		 = 		\Pi_{c_y, l_y}(y) \Pi_{c_z,l_z}(z) \exp[i k^{\star}_z z],  \qquad \quad 			\Pi_{c,l}(s) 	 =   \frac{\tanh[c(s+l)] - \tanh[c(s-l)] }{2}, \\
&  \hat{E}(k_y,k_z) = \frac{\pi^2 }{ c_y c_z}\left(   		\sin[ k_y l_y ]  \csch \left[ \frac{ k_y \pi }{2 c_y } \right] \right) 		\cdot  		\left( \sin[ (k_z^{\star}-k_z) l_z]   \csch \left[ \frac{( k_z^{\star} - k_z ) \pi }{2 c_z} \right] 	\right)	,	
\end{split}
\label{E_wave_maker}
\end{equation}
where the smoothing parameters $c_y$ and $c_z$ must be chosen sufficiently large, as discussed in detail in \S \ref{comparison}, with the discontinuous edges corresponding to $c_y \rightarrow \infty$ and $ c_z \rightarrow \infty$. 
The exact solutions to the linearized equations, constructed in the next section, do not rely on this particular envelope choice, and thus are far more general. 

\begin{figure}
\begin{center}
\begin{minipage}{.45\textwidth}
\hspace*{1cm}
\includegraphics[width =0.6 \textwidth]{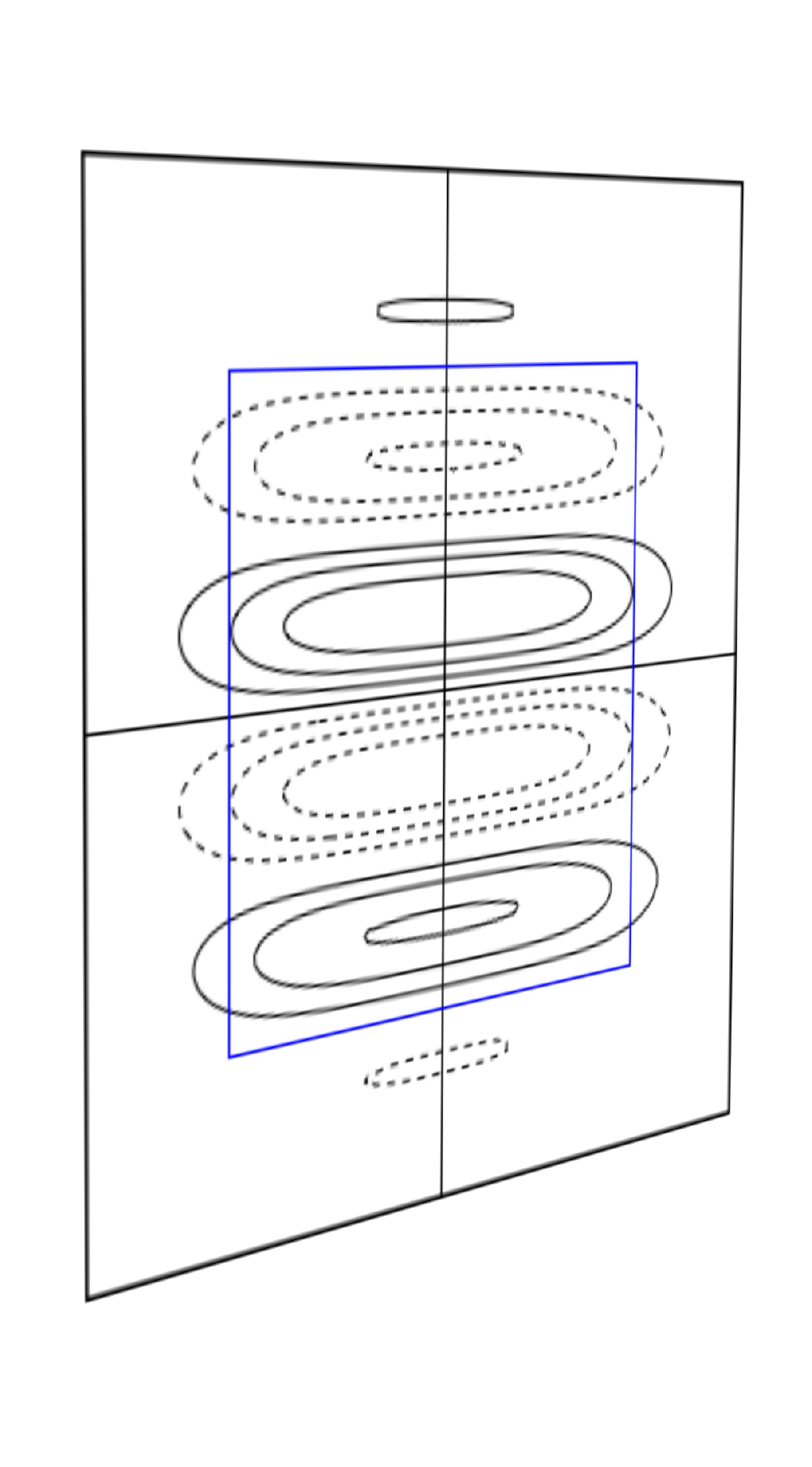}
\put(-105,15){ $ x=0 $}
\put(-120,50){\huge $ \uparrow $}
\put(-125,68){phase}
\put(0,120){$t = 0$}
\put(-63,27){ $ y =0$}
\end{minipage}
\begin{minipage}{.45\textwidth}
\hspace*{1cm}
\includegraphics[width = 0.6 \textwidth]{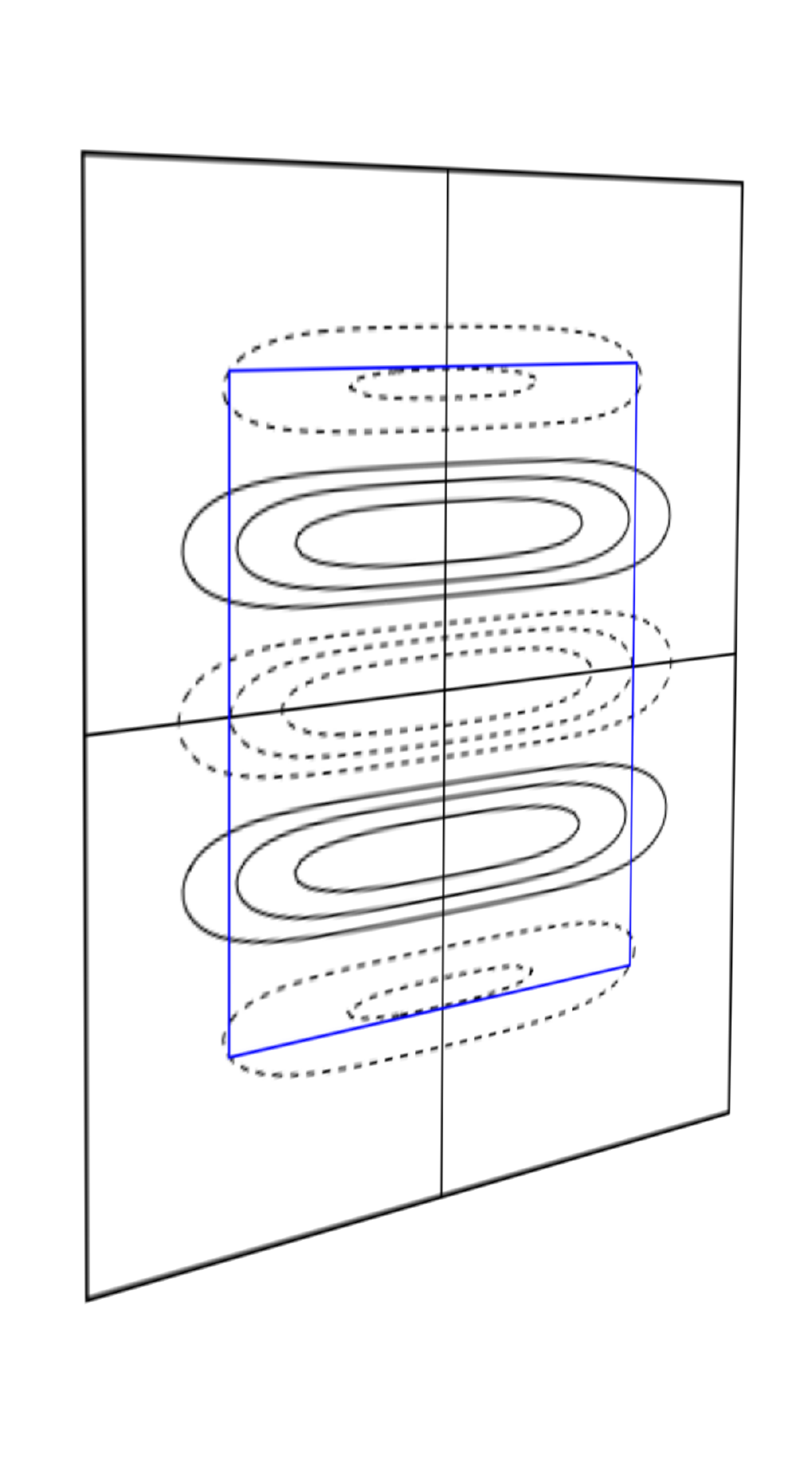}
\put(-105,15){ $ x=0 $}
\put(-120,50){\huge $ \uparrow $}
\put(-125,68){phase}
\put(0,120){$t = \pi/(2\omega)$}
\put(-63,27){ $ y =0$}
\end{minipage}
\caption{Sketch of the imposed boundary oscillation 
in the forcing plane $x=0$, at two instances in time, $t=0$ (left) and $t=\pi/(2\omega)$ (right), for smoothed wave maker envelope given by expression \ref{E_wave_maker} with $c_y=c_z=2.5$, $l_y=l_z=1$, $k_z^{\star} = 2\pi$. The blue box indicates the wave maker domain, $(y,z)\in[-1,1]\times[-1,1]$. The solid and dashed lines show contours of $\dot{a}(y,z,t)$ at $0.25$-intervals.}
\label{Fig_forcing}
\end{center}
\end{figure}
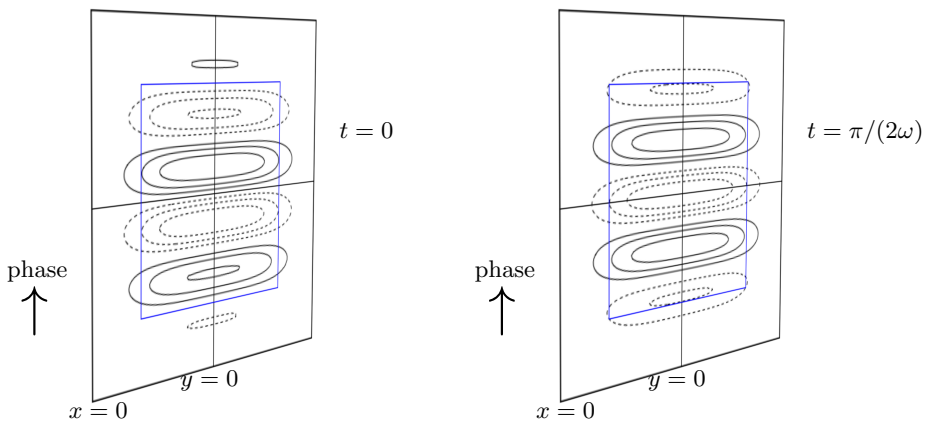

\section{Three-dimensional propagating internal wave beams}
\label{3D_sol}
In this section, we construct analytic expressions for three-dimensional internal wave fields generated by the oscillating boundary constraint, (\ref{constraint_u3}). 
Writing 
$$\hat{W}(x,k_y,k_z) = \int \limits_{-\infty}^{\infty} \int \limits_{-\infty}^{\infty} w(x,y,z,t) \exp[-i k_y y - i k_z z +i \omega t] d y d z ,$$
the governing equation (\ref{eq_w}) reduces to

\begin{equation}
\hat{W}_{xx}	   +    \left(  k_z^2\mu^2  - k_y^2 \right) \hat{W}= 0,
\label{eq_W}
\end{equation}
where
\begin{equation}
\begin{split}
		 \mu^2 	= 	&\frac{   \omega   (-i \omega+\vartheta k^2)^2   + \omega f^2    } {   (  \omega^2 +  i \omega \vartheta k^2 -N^2) ( \omega + i \vartheta k^2 ) }	 =  	\tan^2 \theta		+ 		i \vartheta k^2 \left( \frac{\omega^2 N^2+f^2(N^2 -2\omega^2 )}{\omega (N^2 -\omega^2)^2}\right)	 + 	\mathcal{O}(\vartheta^2), \\ 
   		k 		=	&	 \frac{ k_z }{ \cos \theta }.
\label{mu_etc}
\end{split}
\end{equation}
In deriving Eq. (\ref{eq_W}), we neglected $\mathcal{O}(\vartheta^2)$-terms by approximating the Fourier-transformed Laplace operator with $-k^2  =-\frac{k_z^2}{\cos^2\theta}$ whenever the Laplace operator is associated with viscous dissipation. The validity of this approximation will be apparent from the solution presented below, for which the Fourier-transformed Laplace operator becomes $-(k_x^2 + k_y^2 + k_z^2) = -(\mu^2 +1 )k_z^2 = -k^2 + \mathcal{O}(\vartheta)$, with $\mu$ given in Eq. (\ref{mu_etc}). \\

\begin{figure}
\begin{center}
\hspace*{-1cm}
\includegraphics[width =1 \textwidth]{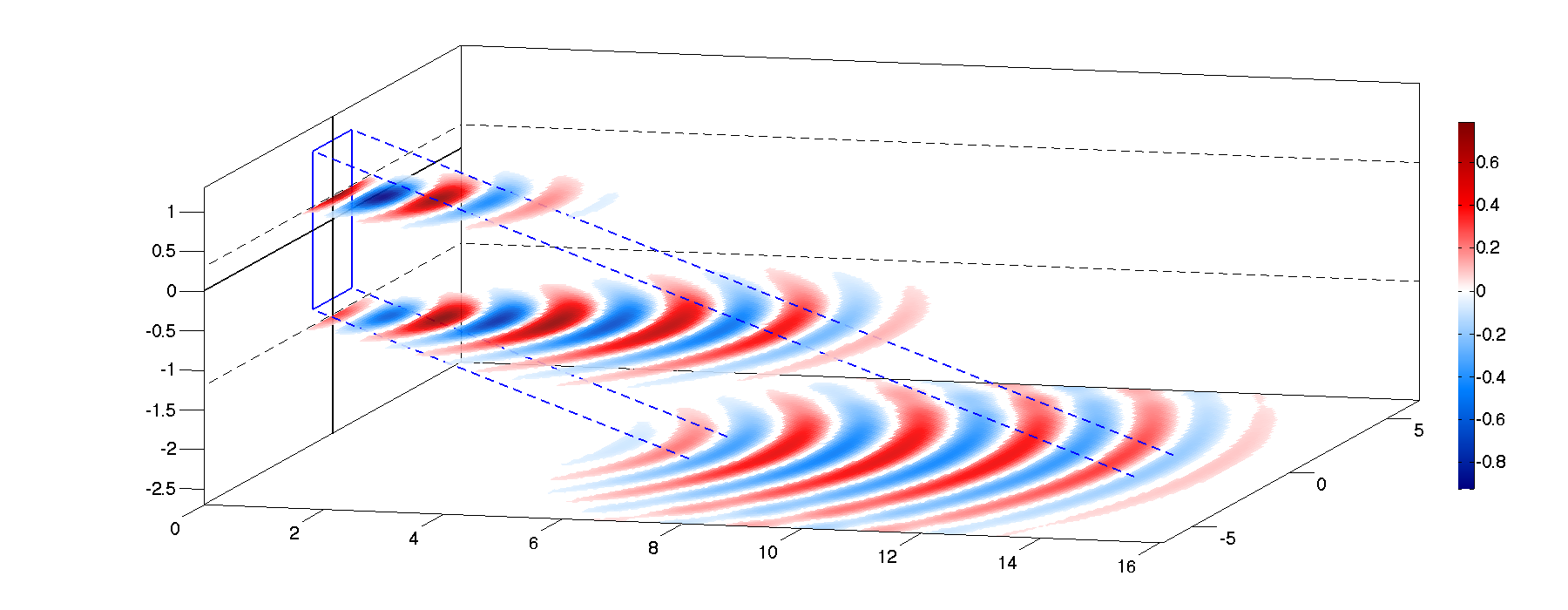}
\put(-220,5){$x/l_y$}
\put(-365,75){$z/l_y$}
\put(-58,25){$y/l_y$}
\put(-15,76){$u^{b}$}
\caption{This figure shows the inviscid ($\vartheta = 0$) non-dimensional $x$-velocity component, $u^{b}$, of a three-dimensional diffracting internal wave beam in three horizontal planes ($z/l_y = 0.3, -1.2$ and $-2.7$), for the smoothed wave maker forcing depicted in Fig. \ref{Fig_forcing}. For clarity, values $|u|< 0.05$ are invisible. The four blue diagonal dashed lines indicate the center wave beam region (set by forcing region $(y/l_y,z/l_y) \in [-1,1]\times [-1,1]$). Due to diffraction, the wave beam widens in the $y$-direction with increasing distance to the source at $x=0$.  }
\label{Fig_3Dsketch}
\end{center}
\end{figure}
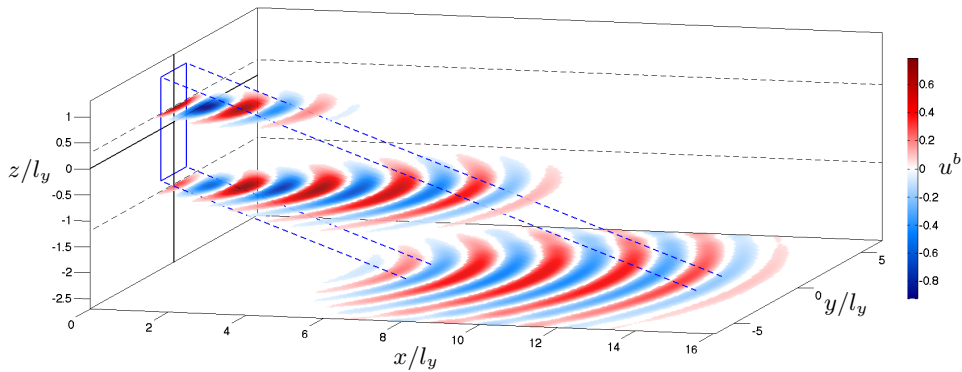

\noindent The homogeneous solution of the 1D Helmholtz equation (\ref{eq_W}), which is bounded for $x>0$, is proportional to $\exp[i k_x x]$, where $k_x = \sqrt{k_z^2 \mu^2 - k_y^2}$.  
This shows that the potential $\phi$ can be written as
\begin{equation}
 \phi(x,y,z,t)	 = \frac{1 }{4\pi^2}  \int\limits_{-\infty}^{\infty} \int\limits_{0}^{\infty} 	\hat{\phi}(k_y,k_z) \exp[ i k_x x + i k_y y + i k_z z - i \omega t]	   d k_y d k_z,
\label{phi_sol}
\end{equation}
with the spectrum $\hat{\phi}(k_y,k_z)$ related to $\hat{W}(x,k_y,k_z)$ (by the continuity equation (\ref{cont_Phi_w})) through
\begin{equation}
    \hat{\phi}(k_y,k_z) = \frac{i}{\mu^2 k_z} \hat{W}(x,k_y,k_z)  \exp[-i k_x x]             	
\label{phi_W}
\end{equation}
Without yet specifying the spectrum $\hat{\phi}(k_y,k_z)$, we can readily write the three-dimensional wave beam velocity field, which we denote by ${\bf u}^b$, in terms of $\phi$ alone (up to $\mathcal{O}(\vartheta)$-accuracy):
\begin{equation}
\begin{split}
{\bf u}^b 
    = \begin{pmatrix}
           u^b \\
           v^b \\
           w
         \end{pmatrix}
         =
         \begin{pmatrix}	
         \partial_x	\\	
	 \partial_y	\\
	- \tan^2 \theta \partial_z	
         \end{pmatrix}
	 \phi +           i \vartheta \beta \begin{pmatrix}	
         0	\\	
	0	\\
	 \partial_{zzz}  
         \end{pmatrix}
	 \phi , 		\qquad 		\beta			  = \left( \frac{\omega^2 N^2 +  f^2(N^2 -2\omega^2)}{  \omega  (N^2 -\omega^2)^2 \cos^2 \theta } \right)
	 \label{beam_velocity}
\end{split}
\end{equation}
The aim is to find a solution approximating the boundary forcing constraint (\ref{constraint_u3}). We start by considering constraint (\ref{constraint_u2}) and approximate $u|_{x=0}$ by $ i k_x^{\star} \phi|_{x=0}$, hence imposing 
$$\hat{\phi}= - \frac{ \omega  }{ k_x^{\star} } \hat{E}(k_y,k_z),$$ 
where $k_x^{\star} = \tan \theta k_z^{\star}$ is the imposed horizontal wave length by the wave maker. While approximating $u|_{x=0}$ by $ i k_x^{\star} \phi|_{x=0}$ does imply a slight violation of the (already approximated) boundary constraint (\ref{constraint_u3}), it happens to come with the benefit of removing an unphysical singularity in $\partial_x v$ at $(x,y)=(0,\pm l_y)$\footnote{We remark that replacing $u|_{x=0}$ by $ i k_x^{\star} \phi|_{x=0}$ also neglects the contribution from the curl-free stream function, $\Psi^{cf}$, determined below. Effectively, replacing $u|_{x=0}$ by $ i k_x^{\star} \phi|_{x=0}$ boils down to approximating $i k_x + |k_y| = i \sqrt{\tan^2 \theta k_z^2 - k_y^2} + |k_y|$ by $i \tan \theta k_z$, valid for all sufficiently small $k_y$, and then replacing $k_z$ by the imposed vertical wave number, $k_z^{\star}$. The sharp spectral peaks at $k_y=0$ and $k_z = k_z^{\star}$ justify these simplifications.}. Avoiding a singularity in $\partial_x v$ at $(x,y)=(0,\pm l_y)$ is essential because $v$ has to be continuous at $(x,y)=(0,\pm l_y)$ in order to vanish at these edges.

Fig. \ref{Fig_3Dsketch} illustrates the spatial structure of the (inviscid) $x$-velocity component, $u^b$, for the smoothed wave maker forcing profile depicted in Fig. \ref{Fig_forcing}. \\
The beam decays in the along-beam direction due to horizontal diffraction at rate $\frac{1}{\cos \theta} \Im [ \sqrt{k_z^2 \tan^2 \theta - k_y^2}]$, which is dominated by those transversal wave numbers $k_y$ that are slightly larger (in absolute value) than $ \mu k_z = \sin \theta k + \mathcal{O}(\vartheta)$. This means that part of the generated internal waves, namely those associated with the wave number pairs $(k_y,k_z)$ satisfying $| k_y/k_z|>\tan \theta $, \emph{cannot} leave the forcing plane, $x=0$ due to diffraction. On the contrary, diffraction is practically absent if the imposed spectrum $\hat{E}(k_y, k_z)$ at $x=0$ practically vanishes for $k_y > k_z \tan \theta $, which is the case for transversally very wide, quasi-2D beams. Interestingly, the diffraction decreases with angle $\theta$, i.e. quasi-horizontal propagating beams diffract much stronger than quasi-vertically propagating beams.  \\
 The viscous attenuation rate per unit distance along the beam, i.e. the real part of $i k_x \cos \theta $ at viscous order $\mathcal{O}(\vartheta)$, is given by $\vartheta k^4 \sin \theta / (2 N  k_x \cos \theta)$. For two-dimensional beams, satisfying $k_x = k \sin \theta + \mathcal{O}(\vartheta)$, we recover the viscous attenuation rate $\vartheta k^3/(2 N \cos \theta)$ \citep{TS73, Li78, Vo03}, which equals $1/(2 \lambda_H \cos \theta)$ in the notation by \cite{Bo12}. 
\\
What remains to be solved is the vorticity equation (\ref{curl_mom}) for the horizontal stream function $\Psi$, such that $v = \partial_y \phi + \partial_x \Psi $ vanishes at the sharp vertical edges of the wave maker, at $(x,y) = (0,\pm l_y)$. Straightforward analysis gives
\begin{equation}
\begin{split}
  			\Psi(x,y,z,t)							= \Psi^{rot}  + s \Psi^{sb}  + (1-s)\Psi^{cf}  , 
\end{split}
\label{Psi_sol}
\end{equation}
consisting of rotational, Stokes boundary layer and curl-free stream functions, with the relative contribution of the latter two determined by the parameter $s$, as discussed below. 
The rotational stream function, 
\begin{equation}
\Psi^{rot} = -\frac{i f} {\omega + i \vartheta k^2} \phi  = \frac{ f}{\omega^2} (-i \omega - \vartheta k^2 ) \phi + \mathcal{O}(\vartheta^2) ,
\end{equation}
 solves Eq. (\ref{curl_mom}) at $\mathcal{O}(\epsilon^0)$ for given $\phi$. In contrast to the other stream function components, this is the only vortex component directly linked to the propagating wave beam ($\phi$), and it may be attributed to the internal wave field. Evidently, the rotational stream function vanishes for non-rotating fluids, to which we restrict the analysis in \S \ref{streaming}.  \\
The Stokes boundary layer stream function, 
\begin{equation}
\begin{split}
\Psi^{sb}  &= \frac{  \omega^2}{4\pi^2 k_x^{\star} }  \int\limits_{-\infty}^{\infty} \int\limits_{0}^{\infty} 	\frac{  k_y \hat{E}{(k_y, k_z)} }{ k_x^{sb}( \omega + i \vartheta k_z^2 )}		\exp[ i k_x^{sb} |x|   +   i k_y y + i k_z z - i \omega t  ]  	     d k_y d k_z ,\\
k_x^{sb}  &= \sqrt{i \frac{ \omega}{\vartheta} - k_y^2 - k_z^2},
\end{split}
\end{equation}
 is the solution of the viscous vertical vorticity equation (\ref{curl_mom}), i.e. 
$$(\partial_t - \vartheta \Delta) \Omega^{sb} =0  , 			\qquad 		\Omega^{sb} = \Delta_h \Psi^{sb} ,$$ 
with $\Omega^{sb}$ the associated vertical vorticity, and satisfying $\partial_x \Psi^{sb}|_{x=0} = - \partial_y \phi|_{x=0} + \mathcal{O}(\vartheta)$. 
Similarly, the curl-free stream function 
 \begin{equation}
 \Psi^{cf} =  \frac{ i \omega  }{4\pi^2 k_x^{\star}}  \int\limits_{-\infty}^{\infty} \int\limits_{0}^{\infty} \mbox{sign}[k_y] \hat{E}{(k_y, k_z)} \exp[ i k_y y- |k_y| x  + i k_z z - i \omega t]	   d k_y d k_z ,
 \label{Psi_cf}
\end{equation}
 is the solution to
$$ \Delta_h \Psi^{cf} = 0 \qquad \mbox{statisfying} \qquad \partial_x \Psi^{cf}|_{x=0} = - \partial_y \phi|_{x=0}. $$
For the wave makers with sharp edges (envelope (\ref{E_wave_maker}) with $c_y = c_z = \infty$), expression (\ref{Psi_cf}) reduces to
\begin{equation}
\begin{split}
& \Psi^{cf} =- \frac{\omega }{2 \pi k_x^{\star} } \left(	\log(r^-) - \log(r^+) \right) \exp[ i k_z^{\star} - i \omega t]		\qquad \quad \mbox{for } |z|<l_z, \\
& 		r^{\pm} = \sqrt{ x^2 + (y\mp l_y )^2} \quad \mbox{ for } \ x>0.
 \label{Psi_cf_sharp_edges}
\end{split}
\end{equation}
Note that this curl-free stream function can be interpreted as originating from two vertical line vortices at $y=\pm l_y$.\\
The undetermined parameter $s$ in Eq. (\ref{Psi_sol}) weighs the relative contribution of the (viscous) Stokes boundary layer and (inviscid) vertical line vortices. 
The $y$-velocity component associated with the Stokes boundary layer, $v^{sb} = \partial_x \Psi^{sb}$, takes amplitude $\mathcal{O}(\vartheta^{-1/2})$ at the vertical sheet, $x=0$, i.e. large for weak viscosity. If we impose a no-slip boundary constraint on $v$ at $x=0$, then $s v^{sb}$ must be balanced by other $\mathcal{O}(1)$-velocity components, implying $s \in \mathcal{O}(\vartheta^{1/2})$. For a free-slip (no-stress) boundary condition on $v$, the parameter $s$ must be zero, because a Stokes boundary layer cannot be established at stress-free boundaries. This illustrates that representing the oscillating boundary by a free-slip or no-slip surface is insignificant for the mean flow generation, allowing us to choose $s=0$ and neglect viscous boundary layer effects for simplicity. \\
From here onwards, we restrict ourselves to non-rotating fluids ($f=0$). Rotational effects on the generation of mean flow are worth to be investigated in a separate paper.


\section{Streaming and inviscid mean flow generation}
\label{streaming}

We are interested in the generation of mean flow, driven by the time-averaged non-linear terms in the governing equations (\ref{ge1_mom})-(\ref{ge1_cont}), known as the mean Reynolds stresses. The mean Reynolds stresses at $\mathcal{O}(\epsilon)$ arise from the products of $\mathcal{O}(\epsilon^0)$-solutions. This process is also referred to as streaming when related to viscous attenuation of the wave field \citep{Li78}, in analogy to acoustic streaming, or as rectification when pertaining to the mean field produced by periodic waves, as in tidal rectification (see \S 6.6 in \cite{GB12} and references therein). \\
The question we want to answer is the following: 
Which wave field components are essential in forcing the potentially energetic vortical induced mean flow? As mentioned earlier, the vortical induced mean flow is associated with mean vertical vorticity, which is the only vorticity component which can accumulate energy in the presence of stratification. Conveniently, the (accelerating) vortical and (non-accelerating) buoyancy advection-induced mean horizontal flow components can be disentangled through a Helmholtz-decomposition,
$$ \bar{u} = \bar{\phi}_x - \bar{\Psi}_y, \qquad \bar{v} = \bar{\phi}_y + \bar{\Psi}_x ,$$
where the over bar denotes time-averaging over one wave period, $T = 2\pi / \omega $. The vertical velocity component, $\bar{w}$, is attributed entirely to the buoyancy advection-induced mean flow. We shortly discuss the buoyancy advection-induced mean flow, before analyzing the generation mechanisms driving the resonantly growing vortical induced mean flow. 

\subsection{Buoyancy advection-induced mean flow}
Balancing the time-independent terms in the buoyancy equation, (\ref{ge1_buoy}), one readily finds at $\mathcal{O}(\epsilon)$ the buoyancy advection-induced mean vertical flow
 \begin{equation}
 \begin{split}
 \bar{w}= & 
 -\frac{\epsilon }{N^2}  \ \overline{ \Re[{\bf u}]\cdot \Re[\nabla b] } 
 \ =\ -\frac{\epsilon }{2N^2} \Re[{\bf u}\cdot \nabla b^*]
 \ =\  - \frac{\epsilon}{2\omega } \Im[{\bf u}  \cdot \nabla w^*], 
 \label{barw}
 \end{split}
 \end{equation}
where $*$ denotes complex conjugate, and we used $b = - i ( N^2/\omega ) w + \mathcal{O}(\epsilon)$. Remarkably, the buoyancy advection-induced mean flow may also be non-zero for 3D \emph{inviscid} wave beams, apparent from substituting (\ref{beam_velocity}) into (\ref{barw}),
\begin{equation}
 \bar{w}   = \frac{\epsilon \tan^2 \theta}{2\omega}\Im[\begin{pmatrix}	
         -\nabla_h	\\	
	 \tan^2\theta \partial_z	\\	
         \end{pmatrix}
	 \phi \cdot \nabla \phi^*_z] +  \mathcal{O}(\vartheta) =   \frac{\epsilon}{2\omega } \Im[ \tan^2 \theta (u u_z^* + v v_z^*) - w w_z^* ]  + \mathcal{O}(\vartheta). 
 \label{barw2}
\end{equation}
This emphasizes the fundamental difference with two-dimensional wave beams, which satisfy $w = \tan \theta u + \mathcal{O}(\vartheta)$, $v=0$, and hence solve the non-linear equations identically in the absence of viscosity \citep{McE73, TA03}. 
\\
In the appendix, \S \ref{Lagrangian_Eulerian_Stokes}, we show that, not surprisingly, this Eulerian vertical mean flow is exactly opposed by the vertical Stokes Drift, such that the net (Lagrangian) vertical particles transport vanishes.  \\
The continuity equation relates the vertical induced mean flow to the horizontal buoyancy advection-induced mean flow, $[\bar{\phi}_x, \bar{\phi}_y]$, 
\begin{equation}
\Delta_h \bar{\phi} = -\bar{w}_z .
\label{poisson}
\end{equation}
For two-dimensional internal wave beams in the $(x,z)$-plane, we recover the 2D induced mean flow, $(\bar{u}, \bar{w} ) = (Q_z, -Q_x)$, with $Q = -\int \bar{w} dx$ the mean stream function, corresponding to Eq. (2.10) of \cite{KC01}\footnote{Note that, confusingly, \cite{KC01} refer to the induced mean flow as Stokes Drift. The apparent factor 4 difference in their expression results from a slightly different definition of the stream function.}, as well as Eq. (3.10) of \cite{TA03}. The mean perturbation on the background stratification, Eq. (2.12) of \cite{KC01} and Eq. (3.11) of \cite{TA03}, is more generally given by 
$$\bar{b} = \epsilon \ \overline{ \Re[{\bf u}]\cdot \Re[\nabla w] } = \frac{\epsilon}{2} \Re \left[ {\bf u} \cdot \nabla w^*   \right]. $$
For a three-dimensional wave beam, the expression (\ref{barw}) for $\bar{w}$ depends on the transversal coordinate $y$, and the 2D Poisson equation (\ref{poisson}) must be solved. The scalar field $\bar{\phi}$ inherits the $\mathcal{O}(\epsilon)$-amplitude from $\bar{w}$, and is thus weak at all times for weakly non-linear internal wave beams. 

\subsection{Vortical induced mean flow}
We now turn to the vortical induced mean flow, $[-\bar{\Psi}_y,\bar{\Psi}_x,0]$, which is confined to the horizontal plane. It evolves slowly over the slow time scale $\tau = \epsilon  t$, governed by time-averaged vertical vorticity equation (\ref{curl_mom}) for $f=0$: 
\begin{equation}
\left( \partial_{\tau} -  \frac{\vartheta}{\epsilon} \Delta \right) \Delta_h \bar{\Psi} 	=   \overline{ R } .
\label{curl_mom_eps}
\end{equation}
As argued in \S \ref{3D_sol}, we can neglect the contribution from a Stokes boundary layer. An additional term $ \frac{1}{\epsilon} f \bar{w}_z	$ appears on the right hand side of (\ref{curl_mom_eps}) if rotation ($f\neq0$) is included, resulting in considerably more complicated dynamics by coupling the buoyancy advection-induced mean flow ($\bar{w}$) with the vortical induced mean flow ($\bar{\Psi}$). 
\subsubsection{Streaming in absence of rotation}
For wave fields in non-rotating fluids ($\Psi^{rot} = 0$), the mean vertical vorticity production resulting from beam-beam interactions associated with viscous dissipation is given by 
\begin{equation}
\begin{split}
\overline{R}^b		& = J\overline{(\Re[w], \Re[\phi_z])   }  = - \vartheta \tilde{\beta} J\overline{ (  \Im [ \phi_{zzz} ],  \Re[ \phi_z ]) }  + \mathcal{O}(\vartheta^2) \\
 				& =  \vartheta  \frac{\tilde{\beta}}{2} \Im \left[ {u^b_z }^* v_{zzz}^b -  {v_z^b}^* u_{zzz}^b  \right]  + \mathcal{O}(\vartheta^2) ,
\label{R_b}
\end{split}
\end{equation}
where $\tilde{\beta} = \tan^2 \theta / (\omega \cos^4 \theta) = \beta$ for $f=0$. Physically, the generation of mean vertical vorticity can be understood as slight tilting of the purely horizontal wave beam vorticity vector by the wave beam velocity field. An illustrative discussion of vortex tilting (also known as vortex twisting) can be found in \cite{Ho97}. \\ 
For internal wave beams with a dominant vertical wave number, say $k_z^{\star}$, we may replace $\partial_z  \rightarrow  i k_z^{\star} $, reducing (\ref{R_b}) to
\begin{equation}
\bar{R}^b = -  \vartheta \tilde{\beta} {k_z^{\star}} ^4 (u^b(T/4) v^b(0) - u^b(0) v^b(T/4) ) + \mathcal{O}(\vartheta^2) ,  
\label{R_b3}
\end{equation}
where $[u^b(t),v^b(t)] = \Re[\nabla_h \phi]$. This expression is particularly useful for laboratory experiments, where $k_z^{\star}$ is imposed by the wave maker, and for which time series of the horizontal velocity field, $[u(t), v(t)]$, in a horizontal plane are typically acquired using PIV\footnote{Particle Image Velocimetry (PIV) is an optical method of flow visualization, which detects the displacement of particles suspended in the fluid. If the particles are sufficiently small (in the sense of negligible particle inertia time scale with respect to the smallest relevant time scale of the experiment), then the particle motion is identical to the fluid parcel motion, and experimental PIV velocity fields correspond to the Lagrangian velocity of the fluid. PIV measurements only represent the Eulerian velocity field if the Stokes Drift is negligible.}. The wave beam velocity field $[u^b(t),v^b(t)]$ can be extracted from $[u(t),v(t)]$ by filtering at the forcing frequency and applying a (discrete) Helmholtz decomposition. Note that expression (\ref{R_b3}) is valid for truly three-dimensional wave fields. For quasi-two-dimensional wave fields, for which $k_x \approx k^{\star} \sin \theta = k_z^{\star} \tan \theta $, valid if $|k_y| \ll  k_z^{\star} \tan \theta $, we can interchange $\partial_z$ with $\cot \theta \partial_x$, reducing (\ref{R_b}) to
\begin{equation}
\bar{R}^b = -  \frac{\vartheta {k^{\star}}^3 }{2 N \cos^2 \theta} \partial_y U^2	 + \mathcal{O}(\vartheta^2)  ,  
\label{R_b2}
\end{equation}
where $U = \sqrt{\phi_x \phi_x^{*}}$ is the $x$-velocity amplitude of the wave beam. This approximate expression is identical with the mean vertical vorticity production term in Eq. (2.16) of \cite{FKA18}. An expression proportional\footnote{The mean vertical vorticity production term in (A18) in \cite{Bo12} is a factor 2 smaller as compared to our expression and the expression by \cite{FKA18}. } to the mean vertical vorticity production  derived by \cite{Bo12} is found upon assuming variations of $U$ to be purely due to viscous attenuation, with rate $\vartheta {k^{\star}}^3/(2N) $ in the $x$-direction, equal to $1/(2 \lambda_H)$ in their notation.\\

Whereas previous derivations \citep{Bo12, KA15, FKA18}, resulting in an expression similar to (\ref{R_b2}) for quasi-two-dimensional wave beams, only stress the importance of horizontal cross-beam variations ($\partial_y U \neq 0$), our more general expression (\ref{R_b3}) links streaming to the elliptical wave motion in the horizontal plane. This new insight implies that streaming is maximized for (nearly) circular wave motion, when $u$ and $v$ are of similar magnitude and out of phase. Strong streaming may thus be expected where two internal wave beams (both propagating upwards or downwards) intersect obliquely, for example two beams with propagation in respectively in $x$ and $y$-direction. Configurations of truly oblique intersections of wave beams, where the angle between the horizontal propagation directions is not 0 or 180 degrees, have not yet been studied, although it seems plausible that oblique interactions of wave beams, generated at and scattered by nearby topographic features, are common in the three-dimensional oceans.

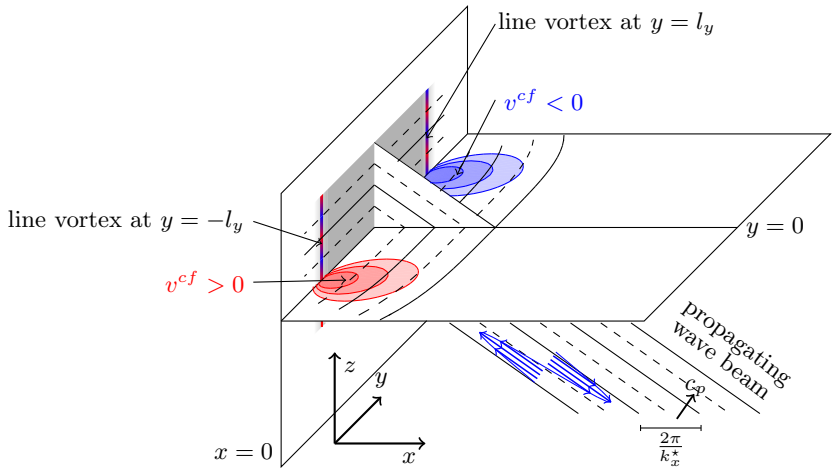
\begin{figure}
\begin{center}
\hspace*{-2.5cm}
\tdplotsetmaincoords{35}{-10}
 \begin{tikzpicture}[scale=0.4]


 \node [rotate=-37] at (12,    -2) {propagating};
 \node [rotate=-37] at (11.5,  -2.5) {wave beam};

\draw[black, thick,->] (1 ,-3,6) -- (4, -3 , 6) node[anchor=north east]{$x$};
\draw[black, thick,->] (1, -3,6) -- (1, 0,   6) node[anchor=north west]{$z$};
\draw[black, thick,->] (1, -3,6) -- (1, -3,  2) node[anchor=south]{$y$};

\draw[->,black,line width=0.25mm] (10 , -4.5 ) -- ({10+ cos(55)*1 },{ -4.5 + sin(55)*1 });
\node[black] at ({10+ cos(55)*1.1 },{ -4.5+sin(55)*1.25 }) {$c_p$}; 

\foreach \d in {2, 4, 6}  
{\draw[-,black] ( 0 , {6 - cot(55)*\d }) -- ({ -\d+ sin(55)*7.35 }, { 6- cos(55)*7.35} ); }
\foreach \d in { 3,  5}  
{\draw[-,dashed,black] (( 0 , {6 - cot(55)*\d }) -- ({ -\d+ sin(55)*7.35 }, { 6- cos(55)*7.35} ); }

\foreach \d in {2, 4, 6, 8}  
{ \draw[-,black]  ({ -\d+ sin(55)*12.8 }, { 6- cos(55)*12.8} ) -- ({ -\d+ sin(55)*18 }, { 6- cos(55)*18} ); }
\foreach \d in { 3,  5, 7}  
{\draw[-,dashed,black] ({ -\d+ sin(55)*12.8 }, { 6- cos(55)*12.8 } ) -- ({ -\d+ sin(55)*18 }, { 6- cos(55)*18} ); }

\node[black,anchor=west] at ( 12, 1.8 ) {$y=0$};
\node[black,anchor=west] at ( -2.5, -2.5, 8 ) {$x=0$};

\draw[black,thin] (0, 1.8, -8) -- (0, 6,-8) -- (0,6,8) -- (0,-3,8) -- (0, -3, -4.5) ;

\fill[gray, opacity=0.6] (0, {6-cot(55)*2}, 4) -- (0, {6-cot(55)*2}, 0)  -- (0, 1.8, 0) -- (0, 1.8, 4) ; 
\fill[gray, opacity=0.6] (0, {6-cot(55)*2}, -4) -- (0, {6-cot(55)*2}, 0)  -- (0, 2.0, -4)  ; 
\shade[left color=gray,opacity=0.6,right color=white] (0, {6-cot(55)*2}, -4) -- (0, {6-cot(55)*2}, -5.2)  -- (0, 1.8, -5.2) -- (0, 1.8, -4.2) -- (0, 2.0, -4) ; 
\shade[right color=gray,opacity=0.6,left color=white] (0, {6-cot(55)*2}, 4) -- (0, {6-cot(55)*2}, 5.2)  -- (0, 1.8, 5.2) -- (0, 1.8, 4) ;
\shade[right color=gray,opacity=0.6,left color=white]  (0, {6-cot(55)*8.2}, 5.2) -- (0, {6-cot(55)*8.2}, 4.15)  --  (0, {6-cot(55)*8.2}, 4) --  (0, {6-cot(55)*8.2}, 4.145)  -- (0, 0.75, 5.2)   ; 

\shade[top color=red,bottom color=blue]  (0, {1.8+ cot(55)*0}, -4.65) -- (0,  {1.8 + cot(55)*1}, -4.65) -- (0,  {1.8 + cot(55)*1}, -4.4) --(0, {1.8+ cot(55)*0}, -4.4)     ;
\shade[top color=blue,bottom color=red]  (0, {1.8+ cot(55)*1}, -4.65) -- (0,  {1.8 + cot(55)*2}, -4.65) -- (0,  {1.8 + cot(55)*2}, -4.4) --(0, {1.8+ cot(55)*1}, -4.4)     ;
\shade[top color=red,bottom color=blue]  (0, {1.8+ cot(55)*2}, -4.65) -- (0,  {1.8 + cot(55)*3}, -4.65) -- (0,  {1.8 + cot(55)*3}, -4.4) --(0, {1.8+ cot(55)*2}, -4.4)     ;
\shade[top color=blue,bottom color=red]  (0, {1.8+ cot(55)*3}, -4.65) -- (0,  {1.8 + cot(55)*4}, -4.65) -- (0,  {1.8 + cot(55)*4}, -4.4) --(0, {1.8+ cot(55)*3}, -4.4)     ;
\shade[top color=blue,bottom color=red]  (0, {1.8+ cot(55)*0},  4.65) -- (0,  {1.8 + cot(55)*1},  4.65) -- (0,  {1.8 + cot(55)*1},  4.4) --(0, {1.8+ cot(55)*0},  4.4)     ;
\shade[top color=red,bottom color=blue]  (0, {1.8+ cot(55)*1},  4.65) -- (0,  {1.8 + cot(55)*2},  4.65) -- (0,  {1.8 + cot(55)*2},  4.4) --(0, {1.8+ cot(55)*1},  4.4)     ;
\shade[top color=blue,bottom color=red]  (0, {1.8+ cot(55)*2},  4.65) -- (0,  {1.8 + cot(55)*3},  4.65) -- (0,  {1.8 + cot(55)*3},  4.4) --(0, {1.8+ cot(55)*2},  4.4)     ;
\shade[top color=red,bottom color=blue]  (0, {1.8+ cot(55)*3},  4.65) -- (0,  {1.8 + cot(55)*4},  4.65) -- (0,  {1.8 + cot(55)*4},  4.4) --(0, {1.8+ cot(55)*3},  4.4)     ;
\draw[red,thick] (0, 0.45, 4.5) -- (0,  {6-cot(55)*8.2}, 4.5) ;  

\draw[black,thin,dashed] (0, {6-cot(55)*3}, -1.5) -- (0,  {6-cot(55)*3}, -6) ;  
\draw[black,thin,dashed] (0, {6-cot(55)*3}, 0) -- (0,  {6-cot(55)*3}, 6) ;  
\draw[black,thin] (0, {6-cot(55)*4}, 6) -- (0,  {6-cot(55)*4}, 0) ;  
\draw[black,thin] (0, {6-cot(55)*4}, -2) -- (0,  {6-cot(55)*4}, -6) ;  
\draw[black,thin,dashed] (0, {6-cot(55)*5}, -3.5) -- (0,  {6-cot(55)*5}, -6) ;  
\draw[black,thin,dashed] (0, {6-cot(55)*5}, 0) -- (0,  {6-cot(55)*5}, 6) ;

\draw[black,thin] (0,1.8, -8) -- (12 ,1.8,-8) -- (12 ,1.8,8) -- (0,1.8,8)  ;  

\draw[black,thin] (12 ,1.8,0) -- (0,1.8,0) ; 
\draw[black, thin] (0, 1.8, 8) -- (0, 1.8, 0) ; 
\draw[black, thin] (0, 1.8, -8) -- (0, 1.8, -4.3) ; 

\foreach \d in {   0.1, 0.2, 0.3, 0.4, 0.5, 0.6, 0.7, 0.8, 0.9, 1.0, 1.1, 1.2, 1.3, 1.4, 1.5, 1.6, 1.7, 1.8}  
{ \draw[-,red,thin] ({  0.55 - 0.6*( cos(180*\d) )    }, { 1.8    }, { 4.5 + 0.7*( sin(180*\d)  ) })  -- ({  0.55 - 0.6* (cos(180*(\d+0.1)) ))  }, {  1.8  }, { 4.5 + 0.7*( sin(180*(\d+0.1)) )  }) ;}
\foreach \d in {   0.1, 0.2, 0.3, 0.4, 0.5, 0.6, 0.7, 0.8, 0.9, 1.0, 1.1, 1.2, 1.3, 1.4, 1.5, 1.6, 1.7, 1.8}  
{ \draw[-,red,thin] ({  1.0 - 1.1*( cos(180*\d) )    }, { 1.8    }, { 4.5 + 1.2*( sin(180*\d)  ) })  -- ({  1.0 - 1.1* (cos(180*(\d+0.1)) ))  }, {  1.8  }, { 4.5 + 1.2*( sin(180*(\d+0.1)) )  }) ;}
\foreach \d in {  0.1, 0.15, 0.2, 0.25,  0.3, 0.35, 0.4, 0.45,  0.5, 0.55, 0.6, 0.65, 0.7, 0.75, 0.8, 0.85, 0.9, 0.95, 1.0, 1.05, 1.1, 1.15, 1.2, 1.25, 1.3, 1.35, 1.4, 1.45, 1.5, 1.55, 1.6, 1.65, 1.7, 1.75, 1.8}  
{ \draw[-,red,thin] ({  1.45 - 1.6*( cos(180*\d) )    }, { 1.8    }, { 4.5 + 1.8*( sin(180*\d)  ) })  -- ({  1.45 - 1.6* (cos(180*(\d+0.05)) ))  }, {  1.8  }, { 4.5 + 1.8*( sin(180*(\d+0.05)) )  }) ;}
\foreach \d in  {   0.1, 0.2, 0.3, 0.4, 0.5, 0.6, 0.7, 0.8, 0.9, 1.0, 1.1, 1.2, 1.3, 1.4, 1.5, 1.6, 1.7, 1.8}  
{ \fill[red, opacity=0.2] ({  0.55 - 0.6*( cos(180*\d) )    }, { 1.8    }, { 4.5 + 0.7*( sin(180*\d)  ) })  -- ({  0.55 - 0.6* (cos(180*(\d+0.1)) ))  }, {  1.8  }, { 4.5 + 0.7*( sin(180*(\d+0.1)) )  }) -- (0, 1.8, 4.5) ; }
\foreach \d in {   0.1, 0.2, 0.3, 0.4, 0.5, 0.6, 0.7, 0.8, 0.9, 1.0, 1.1, 1.2, 1.3, 1.4, 1.5, 1.6, 1.7, 1.8}   
{ \fill[red, opacity=0.2] ({  1.0 - 1.1*( cos(180*\d) )    }, { 1.8    }, { 4.5 + 1.2*( sin(180*\d)  ) })  -- ({  1.0 - 1.1* (cos(180*(\d+0.1)) ))  }, {  1.8  }, { 4.5 + 1.2*( sin(180*(\d+0.1)) )  }) -- (0, 1.8, 4.5); }
\foreach \d in {  0.1, 0.15, 0.2, 0.25,  0.3, 0.35, 0.4, 0.45,  0.5, 0.55, 0.6, 0.65, 0.7, 0.75, 0.8, 0.85, 0.9, 0.95, 1.0, 1.05, 1.1, 1.15, 1.2, 1.25, 1.3, 1.35, 1.4, 1.45, 1.5, 1.55, 1.6, 1.65, 1.7, 1.75, 1.8}  
{ \fill[red, opacity=0.2] ({  1.45 - 1.6*( cos(180*\d) )    }, { 1.8    }, { 4.5 + 1.8*( sin(180*\d)  ) })  -- ({  1.45 - 1.6* (cos(180*(\d+0.05)) ))  }, {  1.8  }, { 4.5 + 1.8*( sin(180*(\d+0.05)) )  })  -- (0, 1.8, 4.5)  ; }

\foreach \d in {   0.4, 0.5, 0.6, 0.7, 0.8, 0.9, 1.0, 1.1, 1.2, 1.3, 1.4, 1.5, 1.6, 1.7, 1.8}  
{ \draw[-,blue,thin] ({  0.55 - 0.6*( cos(180*\d) )    }, { 1.8    }, { -4.5 + 0.7*( sin(180*\d)  ) })  -- ({  0.55 - 0.6* (cos(180*(\d+0.1)) ))  }, {  1.8  }, { -4.5 + 0.7*( sin(180*(\d+0.1)) )  }) ;}
\foreach \d in {     0.5, 0.6, 0.7, 0.8, 0.9, 1.0, 1.1, 1.2, 1.3, 1.4, 1.5, 1.6, 1.7, 1.8}  
{ \draw[-,blue,thin] ({  1.0 - 1.1*( cos(180*\d) )    }, { 1.8    }, { -4.5 + 1.2*( sin(180*\d)  ) })  -- ({  1.0 - 1.1* (cos(180*(\d+0.1)) ))  }, {  1.8  }, { -4.5 + 1.2*( sin(180*(\d+0.1)) )  }) ;}
\foreach \d in {  0.5, 0.55, 0.6, 0.65, 0.7, 0.75, 0.8, 0.85, 0.9, 0.95, 1.0, 1.05, 1.1, 1.15, 1.2, 1.25, 1.3, 1.35, 1.4, 1.45, 1.5, 1.55, 1.6, 1.65, 1.7, 1.75, 1.8}  
{ \draw[-,blue,thin] ({  1.45 - 1.6*( cos(180*\d) )    }, { 1.8    }, { -4.5 + 1.8*( sin(180*\d)  ) })  -- ({  1.45 - 1.6* (cos(180*(\d+0.05)) ))  }, {  1.8  }, { -4.5 + 1.8*( sin(180*(\d+0.05)) )  }) ;}
\foreach \d in  {   0.4, 0.5, 0.6, 0.7, 0.8, 0.9, 1.0, 1.1, 1.2, 1.3, 1.4, 1.5, 1.6, 1.7, 1.8}  
{ \fill[blue, opacity=0.2] ({  0.55 - 0.6*( cos(180*\d) )    }, { 1.8    }, { -4.5 + 0.7*( sin(180*\d)  ) })  -- ({  0.55 - 0.6* (cos(180*(\d+0.1)) ))  }, {  1.8  }, { -4.5 + 0.7*( sin(180*(\d+0.1)) )  }) -- (0, 1.8, -4.3) ; }
\foreach \d in {   0.5, 0.6, 0.7, 0.8, 0.9, 1.0, 1.1, 1.2, 1.3, 1.4, 1.5, 1.6, 1.7, 1.8}   
{ \fill[blue, opacity=0.2] ({  1.0 - 1.1*( cos(180*\d) )    }, { 1.8    }, { -4.5 + 1.2*( sin(180*\d)  ) })  -- ({  1.0 - 1.1* (cos(180*(\d+0.1)) ))  }, {  1.8  }, { -4.5 + 1.2*( sin(180*(\d+0.1)) )  }) -- (0, 1.8, -4.3); }
\foreach \d in {   0.5, 0.55, 0.6, 0.65, 0.7, 0.75, 0.8, 0.85, 0.9, 0.95, 1.0, 1.05, 1.1, 1.15, 1.2, 1.25, 1.3, 1.35, 1.4, 1.45, 1.5, 1.55, 1.6, 1.65, 1.7, 1.75, 1.8}  
{ \fill[blue, opacity=0.2] ({  1.45 - 1.6*( cos(180*\d) )    }, { 1.8    }, {- 4.5 + 1.8*( sin(180*\d)  ) })  -- ({  1.45 - 1.6* (cos(180*(\d+0.05)) ))  }, {  1.8  }, { -4.5 + 1.8*( sin(180*(\d+0.05)) )  })  -- (0, 1.8, -4.3)  ; }

\foreach \d in {  0, 0.5, 1, 1.5, 2, 2.5, 3, 3.5, 4, 4.5, 5, 5.5, 6,6.5}  
{ \draw[-,black,thin] ({  2  -0.0002*\d^4 }, { 1.8    }, { \d   ) })  -- ({  2  -0.0002*(\d+0.5)^4 }, { 1.8    }, { \d +0.5   }) ; }
\foreach \d in {  0, 0.5, 1, 1.5, 2, 2.5, 3, 3.5, 4, 4.5, 5, 5.5, 6, 6.5, 7,7.4}  
{ \draw[-,black,thin] ({  4  -0.0002*\d^4 }, { 1.8    }, { \d   ) })  -- ({  4  -0.0002*(\d+0.5)^4}, { 1.8    }, { \d +0.5   }) ; }
\foreach \d in { 0, 1, 2,3,4,5, 6}  
{ \draw[-,black,thin] ({  1  -0.0002*\d^4 }, { 1.8    }, { \d   ) })  -- ({  1  -0.0002*(\d+0.5)^4 }, { 1.8    }, { \d +0.5   }) ; }
\foreach \d in {  0, 1,2,3,4,5, 6,7}  
{ \draw[-,black,thin] ({  3  -0.0002*\d^4 }, { 1.8    }, { \d   ) })  -- ({  3  -0.0002*(\d+0.5)^4 }, { 1.8    }, { (\d +0.5  ) }) ; }

\foreach \d in {   2.1, 2.5, 3, 3.5, 4, 4.5, 5, 5.5, 6, 6.5}  
{ \draw[-,black,thin] ({  2  -0.0002*\d^4 }, { 1.8    }, { -\d   ) })  -- ({  2  -0.0002*(\d+0.5)^4 }, { 1.8    }, { -(\d +0.5)   }) ; }
\foreach \d in {  0, 0.5, 1, 1.5, 2, 2.5, 3, 3.5, 4, 4.5, 5, 5.5, 6, 6.5, 7,7.4}  
{ \draw[-,black,thin] ({  4  -0.0002*\d^4  }, { 1.8    }, { -\d   ) })  -- ({  4  -0.0002*(\d+0.5)^4 }, { 1.8    }, { -(\d +0.5  ) }) ; }
\foreach \d in { 3.1,4,5, 6}  
{ \draw[-,black,thin] ({  1  -0.0002*\d^4 }, { 1.8    }, { -\d   ) })  -- ({  1  -0.0002*(\d+0.5)^4 }, { 1.8    }, { -(\d +0.5  ) }) ; }
\foreach \d in { 1, 2,3,4,5, 6,7}  
{ \draw[-,black,thin] ({  3  -0.0002*\d^4 }, { 1.8    }, { -\d   ) })  -- ({  3  -0.0002*(\d+0.5)^4 }, { 1.8    }, { -(\d +0.5  ) }) ; }

\draw[->, black,thin] (-4, 2,0) -- (0, 3, 4.5);
\node[anchor=east] at (-4, 2) { \footnotesize line vortex at $ y=-l_y$};
\draw[->, black,thin] (4, 8.5,0) -- (0, 3, -4.5);
\node[anchor=west] at (3.8, 8.5) { \footnotesize line vortex at $ y=l_y$};
\draw[->, black,thin] (4, 6,0) -- (1, 1.8, -4.5);
\node[anchor=west] at (4, 6) { \footnotesize \textcolor{blue}{$v^{cf} <0$} };
\draw[->, black,thin] (-4, 0,0) -- (0.8, 1.8, 4.5);
\node[anchor=east] at (-4, 0) { \textcolor{red}{\footnotesize $v^{cf} >0$} };

\foreach \d in {0, 0.05, 0.1, 0.15, 0.2, 0.25, 0.3, 0.35, 0.4, 0.45, 0.5, 0.55, 0.6, 0.65, 0.7, 0.75, 0.8, 0.85, 0.9, 0.95} 
{ \draw[-,blue] ({ 5.3+ cot(55)*\d - 2*sin(360*(\d   )    )}, { -3.3 + \d + cot(55)*2*sin(360*(\d ))}, 0 )  -- ({ 5.3+ cot(55)*(\d+0.05) - 2*sin(360*(\d  +0.05) )}, { -3.3+ (\d+0.05) + cot(55)*2*sin(360*(\d  +0.05) )} , 0);}
\foreach \d in { 0.125,  0.25, 0.375, 0.625,	0.75, 0.875} 
{ 
\draw[->,blue,line width=0.20mm] ({5.3 + cot(55)*\d }, { -3.3 + \d } , 0) --    ({ 5.3+ cot(55)*\d - 2*sin(360*(\d+0))}, { -3.3 + \d + cot(55)*2*sin(360*(\d+0))}, 0) ; }

\draw[-,thin, black] (8.8, -4.8   ) --  (10.8, -4.8 ); 
\draw[-,thin, black] (8.8, -4.7   ) --  (8.8, -4.9 ); 
\draw[-,thin, black] (10.8, -4.7   ) --  (10.8, -4.9 ); 
\node at (9.8 , -5.5 )  {$\frac{2\pi}{k_x^{\star}}$};

\end{tikzpicture}
\caption{Schematic snapshot of the downward propagating wave beam (solid and dashed phase lines in center plane, $y=0$, and a horizontal plane) and the velocity $v^{cf} = \Psi^{cf}_x$ (red and blue areas), associated with the vertical line vortices at the edges ($y=\pm l_y$) of the wave maker (gray shading at $x=0$ gives $E$, the strength of forcing $F$).  Mean vertical vorticity production through $\bar{R}^{cf}$, dominated by $\frac{1}{2} \Re \left[ w_x^* v_z^{cf} \right]$, occurs primarily in the red and blue shaded areas. The strength of $\bar{R}^{cf} \propto \sin(k_x^{\star} x)/x$ decays radially from the wave maker edges, with sign changes imprinted by the horizontal beam wave number, $k_x^{\star}$. }
\label{fig_induced_mean2}
\end{center}
\end{figure}

\subsubsection{ Inviscid mean flow generation associated with vertical line vortices  }
\label{inviscid_generation}
We now focus on the interaction of the curl-free stream function $\Psi^{cf}$ with the wave beam, possible only through the second wave-vortex interaction term in Eq. (\ref{eq_R}), and given by 
\begin{equation}
\bar{R}^{cf} = -\overline{ \left( \nabla_h   w  \right) \cdot \left( \nabla_h   \Psi^{cf}_z   \right) }  =   \tan^2 \theta \Re \left[ u_z^b v_z^{cf^*}  - v_z^b u_z^{cf^*}  \right]       + \mathcal{O}(\vartheta),  
\label{Rcf}
\end{equation}
where we used $w_x = - u_z  \tan^2 \theta  + \mathcal{O}(\vartheta)$ and $w_y = - v_z\tan^2 \theta + \mathcal{O}(\vartheta)$ for the second expression. This mean vertical vorticity production is particularly interesting, because it may be non-zero even in the absence of viscosity ($\vartheta =0$). For wave beams with dominant vertical wave number, $k_z^{\star}$. we can further simplify (\ref{Rcf}) to
\begin{equation}
\bar{R}^{cf} = \  {k_z^{\star}}^2 \tan^2 \theta \sum \limits_{t \in \{ 0, \frac{T}{4}\} } \left( u^b(t) v^{cf}(t)  - v^b(t) u^{cf}(t) \right)        + \mathcal{O}(\vartheta), 
\label{Rcf2}
\end{equation}
where $[u^{cf}(t),v^{cf}(t)] = \Re[- \Psi^{cf}_y, \Psi^{cf}_x]$. As with (\ref{R_b3}), this expression is also useful if the horizontal velocity field, $[u(t), v(t)]$, is known in a horizontal plane, such as often the case for laboratory experiments. For an experimental data set, it is important to verify whether $[u^{cf}(t),v^{cf}(t)]$, extracted from $\omega$-filtered $[u(t), v(t)]$ through a Helmholtz decomposition, is indeed vertical-vorticity free (within measurement uncertainty).\\
For wave makers with sharp edges at $y=\pm l_y$, such as sketched in Fig. (\ref{fig_induced_mean2}), the velocity field associated with the line vortices decays inversely proportional to the distance to the wave maker edges (see also explicit stream function solution $\Psi^{cf}$, Eq. (\ref{Psi_cf_sharp_edges})). Variations of the wave beam velocity field in a horizontal plane (and within the beam) are dominated by the horizontal beam wave number, $k_x^{\star} = k_z^{\star} \tan \theta$. As a consequence, the mean vertical vorticity production near the edges $y = \pm l_y$, dominated by $\bar{R}^{cf} \approx \frac{1}{2} \Re \left[ w_x^* v_z^{cf} \right]$, is characterized by the sinc function ($\sin(k_x^{\star} x)/(k_x^{\star} x)$), changing sign with increasing distance to the wave maker at rate $\pi/k_x^{\star}$. This is unlike streaming, which does not change sign along the $x$-direction, decaying with distance to the wave maker only due to the (weak) decay of the wave beam strength. This suggests that near the line vortices, inviscid mean flow generation associated with the line vortices prevails, whereas streaming dominates at sufficient distance from the wave maker.

\begin{figure}
\begin{center}
\begin{minipage}{.3\textwidth}
\hspace*{0cm}
\includegraphics[width =1 \textwidth]{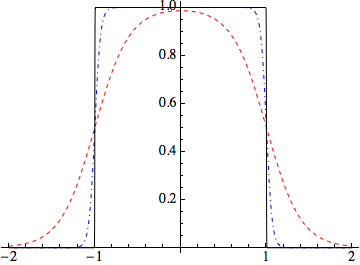}
\put(-10,-6){$y/l_y$}
\put(-33,75){ \tiny	$c= \infty$}
\put(-32,12){ \tiny \textcolor{blue}{$c=\frac{15}{l_y}$} }
\put(-26,26){ \tiny \textcolor{red}{$c=\frac{2.5}{l_y}$} }
\put(-75,88){ $E(y,0) $ }
\put(-110,80){ (a) }
\end{minipage}
\begin{minipage}{.3\textwidth}
\hspace*{0cm}
\includegraphics[width =1 \textwidth]{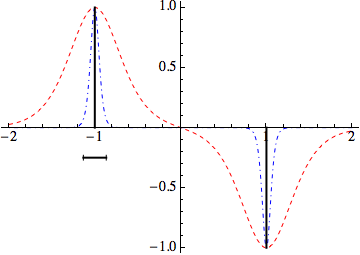}
\put(-15,43){$y/l_y$}
\put(-110,80){ (b) }
\put(-55,78){ $\partial_y E(y,0) $ }
\put(-94,21){ \footnotesize $\frac{2a_0}{l_y}$ }
\end{minipage}
\begin{minipage}{.3\textwidth}
\hspace*{0cm}
\includegraphics[width =1 \textwidth]{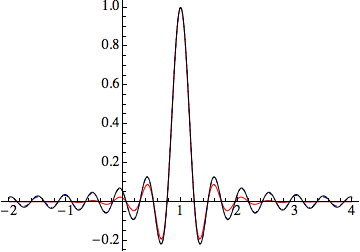}
\put(-22,4){$k_z/k_z^{\star}$}
\put(-110,80){ (c) }
\put(-75,80){ $\hat{E}(0,k_z)$ }
\end{minipage}
\caption{ Presented are in (a) the horizontal wave maker profile $E(y,0)$, defined in Eq. (\ref{E_wave_maker}), and in (b) its derivative, $\partial_y E(y,0)$ (normalized by its maximum),  for $c_y=\infty$ (black, solid), our choice $c_y=15 l_y^{-1}$ (blue, dot dashed), and $c_y=2.5 l_y^{-1}$ (red dashed, choice by \cite{KA15,FKA18}). The vertical lines in (b) illustrate Dirac Deltas, the horizontal line illustrates the relative wave maker excursion, $2a_0/l_y=0.27$, in \cite{Bo12}. (c): The spectrum $\hat{E}(0,k_z)$ with $k_z^{\star} = 3 \pi/l_z$ for $c_z=\infty$ (black, solid), $c_z=15/ l_z$ (blue, dashed dot, almost indistinguishable) and $c_z = 2.5 /l_z$ (red, dashed).}
\label{Fig_env}
\end{center}
\end{figure}
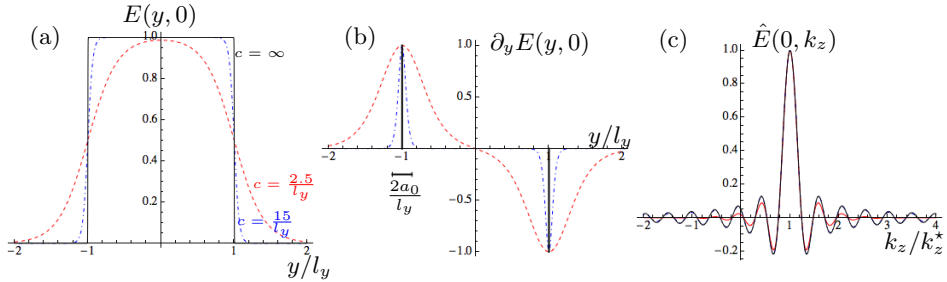


\section{Comparison with \cite{Bo12}}
\label{comparison}
In this section, we compare our theoretical results with the experimental observations by \cite{Bo12}. 
We adopt their parameter values, and specifically the wave frequency $\omega_0 /N_0= 0.26$ and wave maker amplitude $a_0 = 1 $ cm corresponding to the experiments presented in their figures 2 and 3. 
We account for the wave maker inefficiency \citep{Me10} by reducing the wave maker amplitude for the theoretical expressions by 25\%. The associated Stokes number, $\epsilon = 0.2$, is sufficiently small for our perturbational expansion to be valid.   \\
Imperfections in the laboratory set-up justify smooth approximations of the rectangular-shaped wave maker, i.e. finite smoothing parameters ($c_y<\infty$, $c_z<\infty$) in envelope function (\ref{E_wave_maker}). A smooth envelope function $E(y,z)$ is needed to avoid the Gibbs phenomena when numerically integrating the wave beam field, expression (\ref{phi_sol}). We find $c_y = c_z = 15 l_y^{-1}$ justifiable, because $\partial_y E(y,0) \propto \partial_y \Pi_{c_y,l_y}(y)$ then consists of two peaks with widths comparable to the wave maker excursion, $2a_0$, see Fig. \ref{Fig_env}(b). This is not the case for the choice\footnote{Whereas the rectangular-shaped wave maker was horizontally highly smoothed for numerical convenience, \cite{KA15} and \cite{FKA18} did use sharp vertical envelopes, respectively $c_z = \infty$ and $c_z = 7.5/l_z$. It should be noted that they use $Y$ for the vertical and $Z$ for the transverse horizontal coordinate.} $c_y=2.5 l^{-1}$, made by \cite{KA15, FKA18}, also illustrated in Fig. \ref{Fig_env}. 

\begin{figure}
\begin{center}
\hspace*{-0.0cm}\includegraphics[width =1 \textwidth]{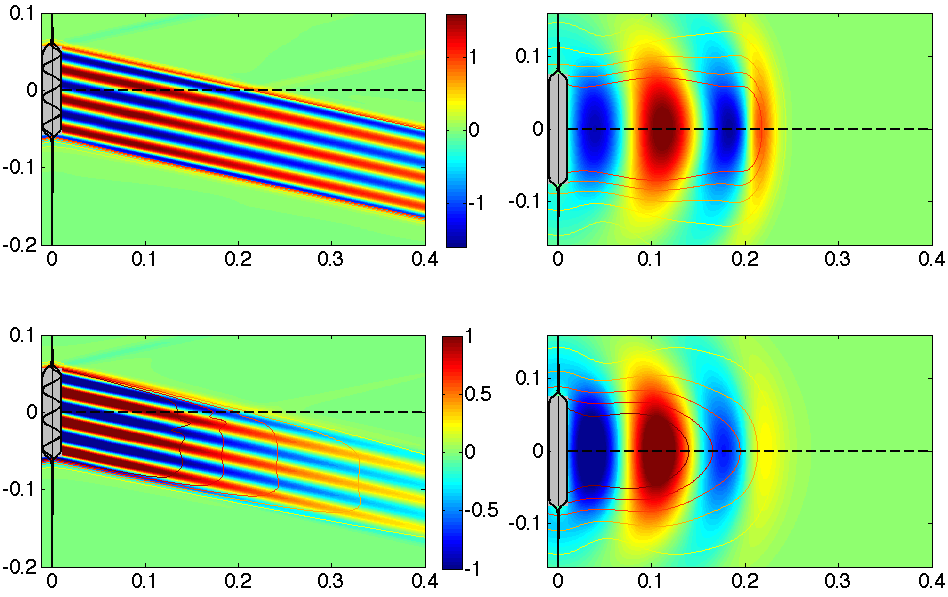}
\put(-300,-5){$x$ [m]}
\put(-300,122){$x$ [m]}
\put(-90,-5){$x$ [m]}
\put(-90,122){$x$ [m]}
\put(-389,60){\rotatebox{90}{$z$ [m]}}
\put(-389,190){\rotatebox{90}{$z$ [m]}}
\put(-180,50){\rotatebox{90}{$y$ [m]}}
\put(-180,180){\rotatebox{90}{$y$ [m]}}
\put(-80,  90){ viscous beam }
\put(-300,90){ viscous beam}
\put(-70,220){ inviscid beam}
\put(-300,220){ inviscid beam}
\put(-392,230){(a)}
\put(-185,230){(b)}
\put(-392,100){(c)}
\put(-185,100){(d)}
\caption{Snapshot of inviscid (a,b) and viscous (c,d) wave field velocity, $u(t) = \Re[u]$ at $t= \pi$, in mm/s, with contours representing the amplitude ($|u|$).  The left panels (a,c) present the side view ($y=0$, dashed line in (b,d)) and the right panels (b,d) present the top view ($z=0$, dashed line in (a,c)). The plots are prepared so as to facilitate direct comparison with the experimental results in figure 2(a,b) of \cite{Bo12}. The idealized wave maker envelope ($c_y = c_z = 15 l_y^{-1}$) is indicated by the black lines. Gray area: extend of oscillating wave maker in laboratory set-up. }
\label{Fig_beam}
\end{center}
\end{figure}

\begin{figure}
\begin{center}
\hspace*{-0.0cm}\includegraphics[width =1\textwidth]{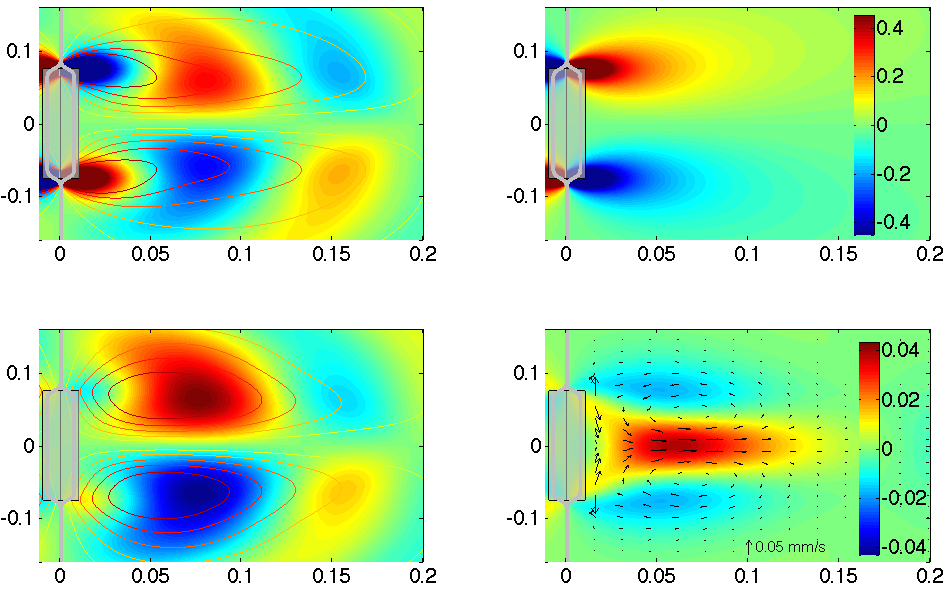}
\put(-300,-5){$x$ [m]}
\put(-300,122){$x$ [m]}
\put(-90,-5){$x$ [m]}
\put(-90,122){$x$ [m]}
\put(-389,50){\rotatebox{90}{$y$ [m]}}
\put(-389,180){\rotatebox{90}{$y$ [m]}}
\put(-185,50){\rotatebox{90}{$y$ [m]}}
\put(-185,180){\rotatebox{90}{$y$ [m]}}
\put(-392,230){(a)}
\put(-185,230){(b)}
\put(-392,100){(c)}
\put(-185,100){(d)}
\caption{Transversal horizontal ($y$-) velocity components in top view ($z=0$) of (a) the wave beam, $v^b$, (b) the line vortices, $v^{cf}$, and (c) their sum, $v$, in mm/s at $t= \pi $, with contours representing their amplitude. All parameter values as in Fig. \ref{Fig_beam}. The wave maker extend (gray area) is transparent, to visualize the singularities of $v^b$ and $v^{cf}$ at the edges, absent for $v$. Plot (d) presents the net Stokes Drift in the $x$-direction, $\bar{u}_S$, with arrows indicating the horizontal Stokes Drift field, $[\bar{u}_S,\bar{v}_S]$, possibly representing the initially observed jet that \cite{Bo12} refer to, as discussed in the text.   }
\label{Fig_v}
\end{center}
\end{figure}

\begin{figure}
\begin{center}
\hspace*{-1cm}
\includegraphics[width =1 \textwidth]{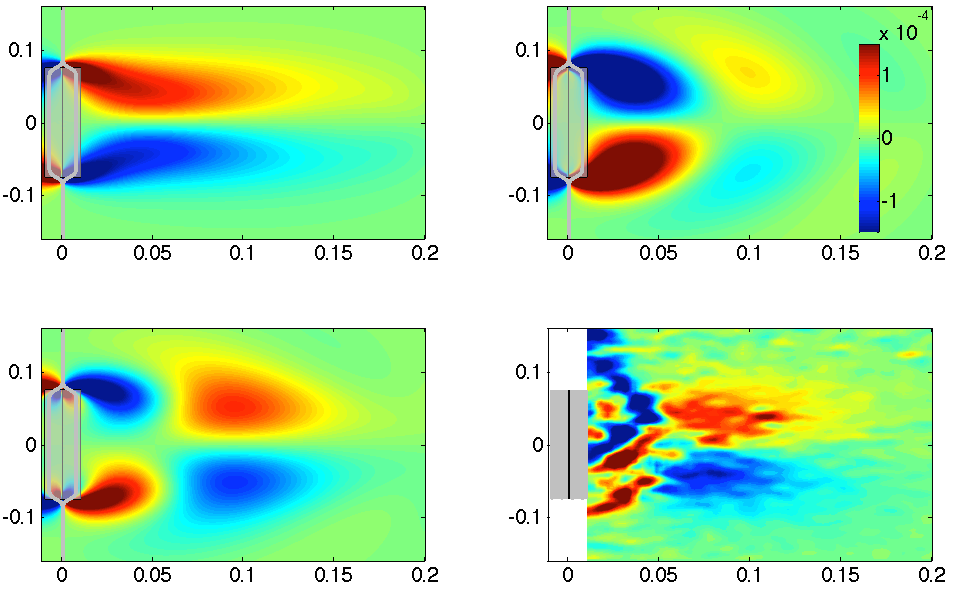}
\put(-280,-5){$x$ [m]}
\put(-100,-5){$x$ [m]}
\put(-389,50){\rotatebox{90}{$y$ [m]}}
\put(-389,180){\rotatebox{90}{$y$ [m]}}
\put(-185,50){\rotatebox{90}{$y$ [m]}}
\put(-185,180){\rotatebox{90}{$y$ [m]}}
\put(-392,230){(a)}
\put(-185,230){(b)}
\put(-392,100){(c)}
\put(-185,100){(d)}
\put(-30,212){ \tiny [1/s$^2$] }
\caption{This figures shows the theoretical mean vertical vorticity production associated with (a) streaming, $\bar{R}^b_0$, (b) interaction of the vertical line vortices with the beam, $\bar{R}^{cf}_0$, and (c) their combined production. The mean vertical vorticity increase, $\partial_t \bar{\Omega}$,  experimentally observed by \cite{Bo12} and presented in their Fig. 3a, is reproduced in (d).    }
\label{U2xy}
\end{center}
\end{figure}

\subsubsection*{Velocity fields of wave beam and line vortices}
We present the $x$-velocity component $u = \phi_x - \Psi^{cf}_y$ in Fig. \ref{Fig_beam}, for an inviscid beam (plots a,b) and a viscous beam (plots c,d), both including the inviscid contribution from the vertical line vortices ($\Psi_y^{cf}$). In agreement with the experimentally observed beam by \cite{Bo12}, their figure 2(a,b), we find that our theoretical viscous beam decays along the propagation direction due to viscous dissipation. The comparison of the viscous and inviscid beams illustrates that viscosity especially smoothes the upper and lower edges of the beam, which feature strong shear in the inviscid case. The intensity of the inviscid beam \emph{also} decays in the along-beam direction. This highlights that a considerable part of the decay in the along-beam direction is due to (inviscid) diffraction, i.e. cross-beam widening. Consequently, the theoretical assumption by \cite{Bo12} that the decay of the beam is purely due to viscous dissipation might be inappropriate for their experimental set-up.  \\
The attentive reader may spot weak upward-propagating wave beams in Fig. \ref{Fig_beam}(a,c), originating from the lower and upper edges of the wave maker. These wave field components, also visible in the experimental data, figure 2(a) of \cite{Bo12}, are associated with the negative part of the spectrum, $\hat{E}(0,k_z)$ for $k_z<0$, visualized in Fig. \ref{Fig_env}(c) and also discussed in \S \ref{slit_exp}. As opposed to the lateral edges of the wave maker, the upper and lower edges function as two line sources of internal waves, generating St. Andrew crosses (also illustrated in Fig. \ref{sketch_energy_input}(a)), with the downward propagating components absorbed in the wave beam (and intensifying the lower and upper beam edges in the inviscid case). \\ 
The beam's $y$-velocity component, $v^b=\phi_y$, the $y$-velocity associated with the line vortices, $v^{cf}=\Psi^{cf}_x$, and their sum, $v$, are presented in Fig. \ref{Fig_v}. Whereas the maxima of $v^b$ and $v^{cf}$ are located at the edges of the wave maker, we find that their sum, $v$, peaks at around $6$ cm from the wave maker. Moreover, $v$ vanishes at $x=0$, consistent with the requirement for the (idealized) wave maker discussed in \S \ref{energy_source}. The maxima of $v^b$ and $v^{cf}$ become singularities for $c_y \rightarrow \infty$, signifying that only their non-singular sum $v$ is physically feasible. This affirms the importance of taking the vertical line vortices into account. 
\\
It is important to realize that the experimental velocity field is observed with PIV, detecting the motion of suspended particles. The particles sizes (about 8 $\mu$m) are sufficiently small such that their motion coincides with the Lagrangian velocity field, representing the Eulerian field only if the Stokes Drift is negligible. The Stokes Drift, discussed in \S \ref{Lagrangian_Eulerian_Stokes}, is secondary to the wave field, but may dominate the time-averaged field, i.e. the observed mean flow. The net theoretical Stokes Drift, $\bar{u}_S$, is plotted in Fig. \ref{Fig_v}(d), exceeding values of 0.05 mm/s, able to transport particles more than 5 cm over the course of the experiment. \cite{Bo12} report that the (Lagrangian) observed mean flow is initially located close to the wave generator. We speculate that they initially observed the Stokes Drift, which appears instantaneously within the beam, dominating the linearly growing (Eulerian) induced mean flow during the first couple of wave periods. Towards the end of the experiment (after $\sim$ 50 wave periods), the Stokes Drift constitutes only 10\% of the Lagrangian mean flow, justifying to treat the Lagrangian PIV observation as an Eulerian field. 
\subsubsection*{Mean vertical vorticity production}
We present the mean vertical vorticity production through streaming ($\bar{R}^b$), associated with the vertical line vortices ($\bar{R}^{cf}$), and their sum ($\bar{R}$) in Fig. \ref{U2xy}(a-c). For comparison, we reproduce the mean vertical vorticity increase observed by \cite{Bo12} in Fig. \ref{U2xy}(d). The similarity of plots (c) and (d) strongly suggests that both viscous streaming and inviscid mean vertical vorticity production associated with the line vortices are significant for this particular experimental set-up. This supports our new hypothesis that inviscid mean flow generation near the edges of wave makers in laboratory set-ups cannot be neglected. It is unclear why the experimentally observed mean vertical vorticity is not symmetric around the center, $y=0$. Our analysis suggests that the dipolar vortex centered around $9$ cm from the wave maker results from streaming, whereas the dipolar vortex closer to the wave maker, at $x=3$ cm, is caused by interaction of the line vortices with the beam. Consistent with the conclusions by \cite{KA15}, the similarity between experimental results with our predictions - excluding mean flow generation associated with beam-modulations - suggests that modulation-effects are not essential for the experiments by \cite{Bo12}.  \\
The mean vertical vorticity associated with the horizontal net Stokes Drift, $[\bar{u}_S, \bar{v}_S]$, instantaneously produced, may be imprinted on the experimental mean vertical vorticity increase if one approximates $\partial_t \bar{\Omega}_L(t) \approx \bar{\Omega}_L(t) /t$, where $\bar{\Omega}_L = \partial_x \bar{v}_L - \partial_y \bar{u}_L$ is the Lagrangian mean vertical vorticity. We verified that this Stokes mean vertical vorticity is only relevant over the first wave period, and insignificant for the experimental data reproduced in Fig \ref{U2xy}(d). \\
It may be argued that the efficiency of the wave maker in generating a wave beam and vertical line vortices differs. If the line vortex generation efficiency is slightly less as compared to the beam generation inefficiency (estimated at 75\%), then we find that the mean vertical vorticity production (not shown) corresponds strikingly well with the experimental mean vertical vorticity increase, Fig. \ref{U2xy}(d). We propose to estimate the strengths of the line vortices experimentally, similar to the study on the efficiency to produce wave beams performed by \cite{Me10}.

\section{Concluding remarks}
\label{conclusions}
Our analysis has once again confirmed that the propagation of three-dimensional internal gravity wave beams differs fundamentally from their two-dimensional counterparts. In agreement with the results by \cite{Bo12} and \cite{KA15}, we find that the finite width of three-dimensional internal gravity wave beams is essential in producing mean vertical vorticity through streaming, resulting in a strong horizontal mean flow. 
\\
Our results are primarily useful for the understanding of laboratory experiments on internal waves, or reversely formulated, for avoiding misinterpretation of experimental results. As such, our work may contribute to correct and insightful extrapolations of experimental results to oceanic circumstances. \\
Importantly, we find that the vertical line vortices at the lateral edges of the wave maker contribute to the mean vertical vorticity production at leading order, and can therefore not be neglected. Moreover, this vertical vorticity production changes sign with increasing distance to the wave maker, the rate being imprinted by the horizontal beam wave length. It is this sign change which results in the quadrupolar structure of the mean vertical vorticity production, which was also observed experimentally. \\
One may wonder why the quadrupolar vertical vorticity production leads to a dipolar induced mean flow (evident in figures 1(b) and 2(d) of \cite{Bo12}), rather than a quadrupolar induced mean flow. The answer is surprisingly simple. The stress of the wave maker on along-boundary mean flow produces a boundary layer with thickness $\sqrt{\nu t}$ (see e.g. \cite{SG00}), reaching $1$ cm thickness within 4 wave periods and almost 4 cm by the end of the experiment. This means that the dipole associated with the vertical line vortices eventually ends up inside the boundary layer of the vortical induced mean flow, and is thus strongly damped by wall friction. The good agreement of the simulated induced mean flow by \cite{KA15} with the experiments stems from the circumstance that the near-field induced mean flow (associated with the wave-line vortex interaction and ignored in their theory) quickly reaches a state in which its energy input matches the wall-friction dissipation rate, whereas the far-field induced mean flow (forced by streaming) accumulates mean vertical vorticity throughout the entire experiment.
\\
A standard paradigm (also known as non-acceleration theorem) widely used in fluid dynamics says that the resonantly growing induced mean flow can only arise when and where waves are (A) dissipated or (B) generated (e.g. \cite{AM78}). In our setting, streaming belongs to the dissipative processes (A), which may occur anywhere in space, whereas the inviscid generation mechanism associated with the vertical line vortices only occurs in the vicinity of the energy source (B). The notion of `vicinity' is obviously problem-dependent, ranging from few centimeters in laboratory set-ups to possibly dozens of kilometers in oceanographic settings. Although counterexamples are known \citep{BM05}, this paradigm nevertheless forms a suitable conceptual classification of the two vortical mean-flow generation mechanisms discussed here.  \\
The Helmholtz decomposition has proven extremely useful in many studies on fluid dynamics (see e.g. the review by \cite{Bh13}), including recent developments for internal wave data analysis \citep{Bu17}. Once more, we find that disentangling the wave horizontal wave field with a Helmholtz decomposition into propagating internal wave ($\phi$) and non-propagating oscillation ($\Psi$) is essential in determining the mean vertical vorticity production contributions. It is now a standard procedure to disentangle experimental wave field data in \emph{vertical} planes into field components propagating in different vertical directions through Hilbert filtering \citep{Me08}. We propose to extend the experimental wave field decomposition procedure with a (discrete) Helmholtz decomposition applied to the horizontal velocity field in \emph{horizontal} planes to disentangle wave and vortex components. 
\\
Possibly most importantly, our analysis in \S \ref{energy_source} illustrates that an appropriate mathematical representation of the wave maker is essential in studies on mean flow generation. The numerical code by \cite{Si16}, used in several studies involving wave makers, and the numerical code employed by \cite{Gr13, Ra18} for simulations of mean flow generation upon reflection at inclined bottoms, do not capture the vertical line vortices at the edges of the wave makers. While the absence of the vertical line vortices in simulations may be of no concern for the far field, one must be aware that the vertical line vortices are intrinsic to laboratory experiments, and may impact the non-linear dynamics in the vicinity of the wave maker. \cite{Br16} find mean flow generation in their numerical simulations, possibly related to streaming in the boundary layer at the rigid walls. A similar study by \cite{Pi18} places a wave maker in a much wider tank, resulting in lateral spreading of the internal wave field. Based on our analysis, we expect mean flow generation associated with the line vortices at the edges of the wave maker in the laboratory, this being absent in the corresponding numerical simulation. The strengths of our theoretical study is that we can 'switch' on and off the vertical line vortices, thereby investigating whether incorporating the line vortices may be important for more complicated configurations that can be tackled only numerically. 
\\
This work also forms the basis for further research on the effect of rotation ($f\neq 0$) on strong mean flow generation. Numerical simulations by \cite{Ra18}, as well as recent work by \cite{WY15, FKA18}, indicate mean flow generation to be strongly influenced by rotation. Using the wave maker representation derived in \S \ref{energy_source}, it is straightforward to incorporate the effect of rotation on streaming, to be presented in a separate paper soon.


\section*{Acknowledgement}
We thank T. R. Akylas, T. Dauxois, J.-B. Flor, J. Frank, G.-J. van Heijst, S. Joubaud, A. M. S. Kruseman, J. Sommeria, C. Staquet, A. Venaille and B. Voisin for helpful discussions. Suggestions by three anonymous reviewers led to a significantly improved manuscript. F.B. and K.R. are thankful to the organizers of the FDSE summer school 2016 in Cambridge, UK, which initiated the collaboration leading to this article. F.B. is grateful for support by NWO Mathematics of Planet Earth grant 657.014.006.

\section{Appendix}
\subsection{(Lagrangian) induced particle transport}
\label{Lagrangian_Eulerian_Stokes}
The motion of fluid parcels is described in the Lagrangian framework, which follows a parcel as it moves through space and time.  Experimental mean flow fields derived with PIV correspond to the Lagrangian mean flow. This section shortly discusses the relation between Lagrangian and Eulerian wave fields, which is needed for the comparison of our theory (in Eulerian framework) with the experimental results (in Lagrangian framework) by \cite{Bo12} in \S \ref{comparison}.
We denote field variables in the Lagrangian framework with subscript $L$, i.e. $u_L$ for the $x$-velocity component. The Lagrangian velocity field ${\bf u}_L$ is related to the Eulerian velocity, ${\bf u}_E$, through
\begin{equation}
{\bf u}_L = {\bf u}_E + {\bf u}_S, 
\label{u_L}
\end{equation}
where ${\bf u}_S$ is the so-called Stokes Drift (see e.g. \cite{Bu10}), and ${\bf u }$ may be replaced by any field variable\footnote{If ${\bf u}$ in Eq. (\ref{u_L}) is replaced by a field variable other than a velocity component, i.e. by the buoyancy $b$, then $b_S$ is referred to as the Stokes \emph{Correction}.}. 
 The description of the Stokes Drift, which is in fact \emph{defined} by Eq. (\ref{u_L}), is in general nontrivial. Mean field quantities in the Lagrangian framework are in the most general setting described by the Generalized Lagrangian mean theory, developed by \cite{AM78}. For sufficiently small Stokes number ($\epsilon \ll 1$) and over sufficiently short times $t-t_0$, we can conveniently express the Stokes Drift at position ${\bf x}_0$ for times $t>t_0$ as
\begin{equation}
{\bf u}_S({x}_0,t) = \epsilon  \left( {\bf x}(t) \cdot {\bf \nabla}   \right) {\bf u}_E	+ \mathcal{O}(\epsilon^2) 
\label{u_S}
\end{equation}
where 	
$${\bf x}(t) 	= 	    \int_{t_0}^t {\bf u}_E({\bf x}(t'), t') dt'  		= 	     \int_{t_0}^t {\bf u}_E({\bf x}_0, t') dt'  		+ 		\mathcal{O}(\epsilon) $$
is the fluid parcel displacement at time $t$ with respect to the initial particle position ${\bf x}(t_0) = {\bf x}_0$ at time $t=t_0$. The small Stokes number $\epsilon=\frac{U_0}{L_0 \omega_0} \ll 1$ appears in (\ref{u_S}) because ${\bf u}_S$ is non-dimensionalized with $U_0 = \epsilon L_0 \omega_0$, while ${\bf x}(t)$ is non-dimensionalized by $U_0/\omega_0$. The Stokes Drift averaged over one wave period, $\bar{{\bf u}}_S$, can thus be expressed at $\mathcal{O}(\epsilon)$-accuracy as
\begin{equation}
 \bar{{\bf u}}_S = \epsilon \overline{  \left(     \int_{t_0}^t {\bf u}_E({\bf x}_0, t') dt'  \cdot {\bf \nabla}  \right)    {\bf u}_E  }    =    \frac{\epsilon}{2\omega} \Im \left[ \left( {\bf u} \cdot {\bf \nabla} \right) {\bf u}^* \right], 
 \label{u_S2}
\end{equation}
assuming that the time average of ${\bf u}_E = \Re[ {\bf u}]$ vanishes at leading order $\mathcal{O}(1)$, i.e. $\bar{u} = 0 + \mathcal{O}(\epsilon)$. The (theoretical) horizontal Stokes Drift components, $[\bar{u}_S, \bar{v}_S]$, are presented in Fig. \ref{Fig_v}d for the experimental parameter values by \cite{Bo12}. We find that the vertical Stokes Drift, $\bar{w}_S$, is identical in magnitude and opposite in sign to the vertical induced mean flow, $\bar{w}_E = \bar{w} $ in Eq. (\ref{barw}), such that 
\begin{equation}
 \bar{w}_L = \bar{w}_E + \bar{w}_S = 0 
\label{barwL}
\end{equation}
up to $\mathcal{O}(\epsilon)$-accuracy.  This means that internal waves cannot transport mass vertically through streaming, a result which has been well-known for a long time for monochromatic 2D internal waves \citep{Wu71, OM86}, and may be expected based on the conservation of the prescribed stratification. To the best of our knowledge, we are the first to explicitly verify this for monochromatic 3D internal waves. Our analysis can easily be extended to time-periodic internal waves (i.e. superposition of monochromatic beams whose frequency ratios are rational). \\
In 2D, the absence of vertical Lagrangian mean flow, Eq. (\ref{barwL}), together with mass conservation (continuity equation) implies the absence of Lagrangian mean flow altogether. This is not necessarily the case for 3D internal wave beams, stressing once again the importance of considering truly three-dimensional internal wave configurations. \\
The numerical investigation by \cite{Bi97} predicts wave-induced chaotic mixing of particles (fluid parcels) for the superposition of at least three monochromatic inviscid 3D internal waves. Their analysis neglects the generation of (Eulerian) mean flow, which oversimplifies matters as shown by our perturbational analysis; the absence of net vertical particle transport evidently inhibits any particle mixing in the vertical direction. Nevertheless, wave-induced chaotic particle mixing may still occur in the horizontal plane, even for 3D \emph{inviscid} internal wave beams.

\subsection{Internal wave slit-experiment}
\label{slit_exp}

As a side remark, we want to mention that the non-vanishing of the wave maker envelope spectrum $\hat{E}(k_y,k_z)$ for $k_z<0$ in expression (\ref{E_wave_maker}) may explain an unsolved diffraction problem by \cite{Me08}. Using a similar laboratory set-up as \cite{Bo12}, \cite{Me08} generate an upward-propagating internal wave beam impinging onto a slit. Their slit-experiments reveal that the transmitted internal waves propagate upward \emph{and} downwards; the relative strengths of the downward propagating wave increases with decreasing slit-height, $s_0$. The downward propagation may appear surprising, because classical ray theory (e.g. \cite{Li78}) predicts that all transmitted internal waves should continue propagating upwards. No theoretical explanation of the experiment has yet been provided. \\
We claim that the explanation of the slit-diffraction-problem is surprisingly simple. The width of the spectral peak at $k_z = k_z^{\star}$ in Fig. \ref{Fig_env}c increases with decreasing wave beam height, $2l_z$. For sufficiently large $l_z$, such as  in Fig. \ref{Fig_env}c and for the impinging internal wave beam in \cite{Me08}, the peak width is relatively thin, and wave propagation is predominately upwards (for peak at $k_z = k_z^{\star}$) or downwards (for peak at $k_z = -k_z^{\star}$). For the transmitted internal waves in the slit experiment, one must replace $l_z$ in expression (\ref{E_wave_maker}) by $s_0/2$, half of the slit height. For the smallest slit heights used by \cite{Me08}, the width of the peak at $k_z = -k_z^{\star}$ (not shown) becomes much wider than $1$, hence the transmitted wave energy is split almost equally into upward and downward propagating components.

\bibliographystyle{jfm}
\bibliography{diffracting_wave_beam}

\end{document}